\documentclass[reprint,amsmath,amssymb,aps,pra,floatfix]{revtex4-2}

\usepackage[T1]{fontenc}
\usepackage{lmodern}
\usepackage{amsmath,amssymb,bm}
\usepackage{booktabs}
\usepackage{graphicx}
\usepackage{xcolor}
\usepackage{hyperref}
\usepackage{placeins}
\usepackage{enumitem}
\usepackage{float}

\begin{document}

\title{Symmetry Adapted Hierarchical Equations of Motion for Exact Simulations of Large Polariton Systems}

\author{M. Elious Mondal}
\email{mmondal@ur.rochester.edu}
\affiliation{Department of Chemistry, University of Rochester, Rochester, New York 14627, USA}
\author{Pengfei Huo}
\email{pengfei.huo@rochester.edu}
\affiliation{Department of Chemistry, University of Rochester, Rochester, New York 14627, USA}
\affiliation{The Institute of Optics, Hajim School of Engineering, University of Rochester, Rochester, New York 14627, USA}

\begin{abstract}
\noindent Hierarchical equations of motion (HEOM) provide exact dynamics of open quantum systems coupled to harmonic baths, but their computational cost becomes prohibitive for systems with many independent local environments. In this work, we develop a symmetry-adapted HEOM formalism to significantly reduce the computational cost for the permutationally invariant Holstein-Tavis-Cummings (HTC) model. The method removes redundant information in two stages. First, all auxiliary density operators (ADOs) related only by relabeling identical molecules and their bath channels are replaced by a single canonical occupation-pattern representative. Second, molecules with the same local hierarchy occupation produce repeated matrix elements within each representative, allowing only the distinct complex variables to be propagated instead of the full $(N+1)\times(N+1)$ ADO matrices. The resulting matrix-free equations are evaluated using precomputed connections and molecular multiplicities. At fixed hierarchy depth $L$ and number of bath correlation exponentials $m$, the number of canonical representatives becomes independent of the ensemble size for $N\geq L$ and the number of unique variables saturates for $N\geq L+2$. The formulation easily extends to multiple-exponential bath decompositions, arbitrary initial density operators, static disorders, and cavity loss. Our benchmarks reproduce conventional HEOM dynamics while requiring far fewer propagated variables and substantially less memory.
\end{abstract}
\maketitle

\section{Introduction}

Collective strong light-matter coupling hybridizes molecular excitations with a confined photon mode to form polaritonic states distributed across a molecular ensemble.~\cite{Mandal2023CR,Ying2026ARPC} The Holstein-Tavis-Cummings (HTC) model provides a widely used microscopic description of this setting: the cavity couples collectively to the molecular transitions, whereas vibrational and environmental fluctuations generally remain local to individual molecules. These local environments drive relaxation, dephasing, and population exchange between bright, dark, and photonic sectors, and their memory can be important on the timescales of polariton dynamics~\cite{Chng2024JPCL,Hu2025JCP,Mondal2026PNAS}. Accurately treating such non-Markovian dynamics therefore requires a method that retains molecularly resolved environmental fluctuations while remaining computationally practical as the number of molecules increases.

Hierarchical equations of motion (HEOM)~\cite{Tanimura1989JPSJ, Ishizaki2005JPSJ, Tanimura2006JPSJ, Tanimura2020JCP} provide a nonperturbative and systematically convergent description of harmonic environments once their correlation functions are represented by finite sums of exponential terms~\cite{Ishizaki2005JPSJ,Bai2024ACR,Lambert2023PRR}. Their direct application to a molecular ensemble, however, encounters a severe dimensional bottleneck. For $N$ molecules, $m$ exponential terms per local bath, and total hierarchy depth $L$, conventional HEOM contains
\begin{subequations}
\begin{equation}\label{eq:intro-direct-ado-count}
\mathcal N_{\mathrm{ADO}}(N,m,L)
=
\binom{Nm+L}{L}\notag    
\end{equation}
Auxiliary density operators (ADO) and a total of
\begin{equation}\label{eq:intro-direct-variable-count}
\mathcal{N}_{\mathrm{var}}(N,m,L)
=
\mathcal N_{\mathrm{ADO}}(N,m,L)(N+1)^2 \notag      
\end{equation}
\end{subequations}
complex variables when every ADO is stored as a dense matrix in the single excitation basis. For identical molecules, much of this growth represents redundant information. Conventional HEOM separately retains every molecular assignment of the same local hierarchy occupations and stores many matrix elements that are repeated because the associated molecules are physically interchangeable.

Several complementary developments have improved the efficiency of HEOM, including scaled ADOs~\cite{Shi2009JCP, Zhu2011JPCB}, filtering of negligible hierarchy elements~\cite{Zhang2017JCP, Temen2020IJQC,Huang2023CP}, compact bath correlation decompositions~\cite{Liu2014JCP, Chen2022JCP}, parallel propagation~\cite{Kreisbeck2011JCTC, Tsuchimoto2015JCTC, Kramer2018JCC, Zhang2024WIRCMS}, and low-rank or tensor representations~\cite{Shi2009JCP, Zhu2011JPCB, Liu2014JCP, lindoy2019new, yan2021JCP, Lambert2023PRR, ke2023JCP, Bai2024ACR, lindoy2025JCP, rodriguez2026JCTC, chen2025JCP, li2026arXiv, mangaud2023survey, guan2024JCP}. Permutationally invariant formulations have also reduced the computational cost of quantum trajectories~\cite{Lloyd2026arXiv}, while collective structure and Hamiltonian sparsity have been exploited in polariton simulations~\cite{Mondal2025JCP1, Mondal2025JCP2, Chng2025NL, Chng2026NL, koshkaki2026NC}. Mean-field dynamical approaches have been employed to exactly evaluate observables such as photon populations and linear spectra in the thermodynamic limit ($N\rightarrow\infty$)~\cite{li2025JCP, mu2026PRL, fowler2022PRL, fowler2023PRR, fowler2024Thesis, fux2024JCP, ying2026JPCL, perez2023PNAS, perez2025JCP, yuen2024JCP, lindoy2024Nph}. However, these methods are not applicable for determining polariton or localized populations, nor can they be reliably used for finite, small values of $N$. These approaches do not directly remove the sources of redundancy present in molecule-resolved HEOM. Reducing only the system Hilbert space is insufficient because independent local environments preserve molecule-specific bath histories and mediate dynamics between collective bright states and the molecular dark state sector. An efficient HEOM treatment must therefore compress the hierarchy labels and the internal ADO representation together.
\begin{figure*}
    \centering
    \includegraphics[width=\linewidth]{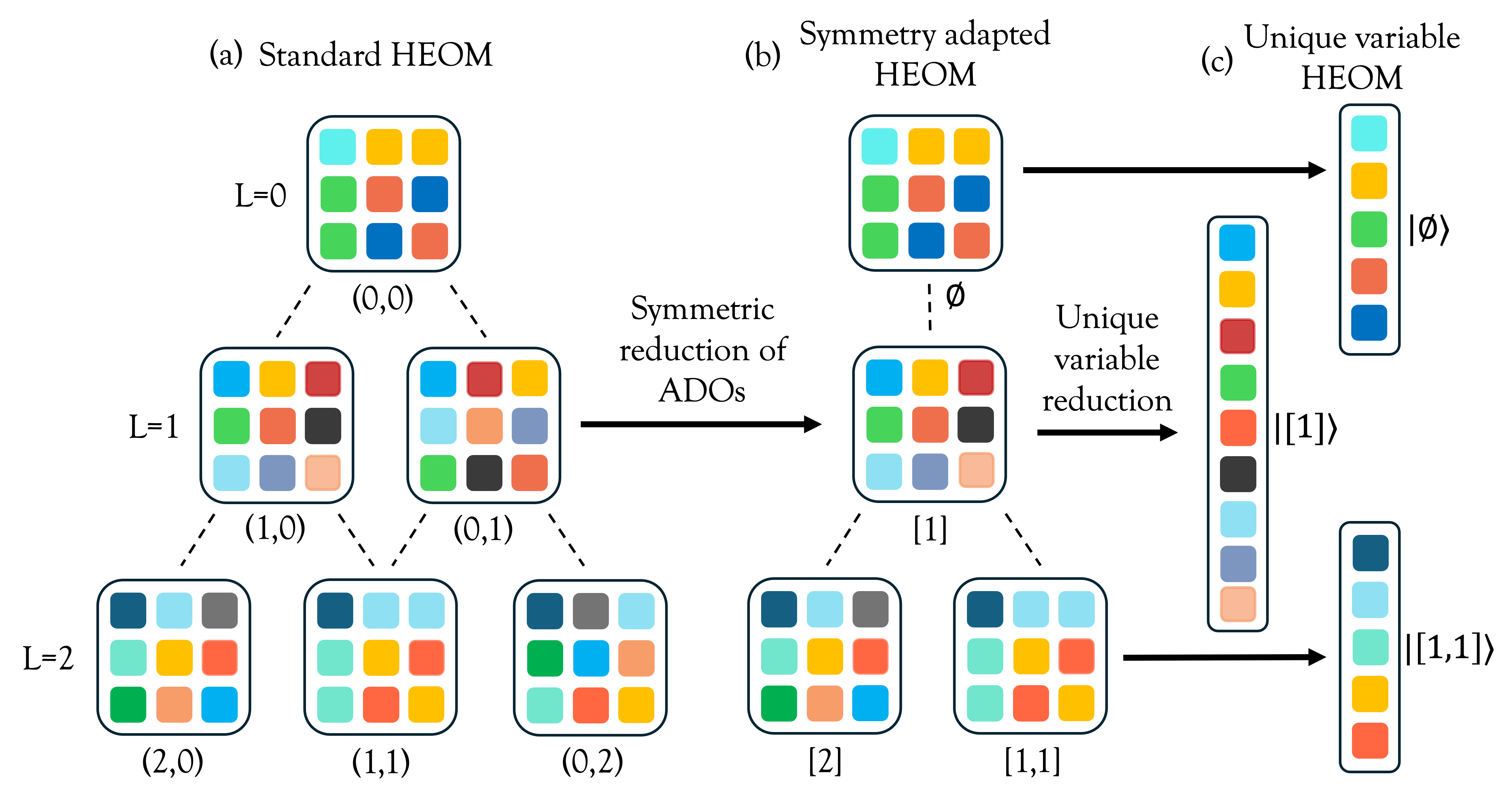}
    \caption{
    A schematic representation of the symmetry adaptation of HEOM for the HTC system. (a) is the ADO structure for the conventional HEOM hierarchy. (b) is the reduction of ADOs within different layers to representatives due to permutation symmetry. (c) is the further reduction of the ADO representatives to their unique variables.
    }
    \label{fig:ADO_count_normal_vs_sym}
\end{figure*}

In this work, we develop a two-stage symmetry-adapted HEOM construction for identical HTC ensembles. First, ADOs related by simultaneous relabeling of molecules and their associated local hierarchy occupations are replaced by one canonical occupation-pattern representative. Second, molecules carrying the same local hierarchy occupation are grouped into occupation categories, and only the resulting unique matrix elements are propagated. Canonical raising and lowering connections, molecular multiplicities, grouped Hamiltonian sums, and aligned target accessors are precomputed and assembled into a matrix-free right-hand side, so neither full ADO matrices nor a dense HEOM generator is constructed during propagation. This procedure is an exact algebraic reduction of the same finite HEOM and introduces no approximation beyond the chosen bath correlation decomposition, hierarchy closure, system Hamiltonian, and numerical integrator. At fixed $m$ and $L$, the number of canonical ADO representatives saturates for $N\geq L$, and the fully compressed unique variable dimension saturates for $N\geq L+2$. This saturation concerns computational dimension, not physical convergence with ensemble size. The equations can retain an explicit dependence on $N$ through molecular multiplicities and through the single-molecule coupling $g$, including calculations in which the collective Rabi splitting is held fixed. The formalism can be easily extended to systems beyond the HTC Hamiltonian as long as they have the permutation relabelling symmetry of the different system subcomponents. The method can also be easily adapted to account for static inhomogeneous diagonal and off-diagonal coupling disorders.

We validate the reduced propagation against conventional QuTiP HEOM~\cite{Lambert2023PRR, Lambert2026PR} calculations for the accessible small ensembles and use the existing calculations to examine three classes of initial conditions: a permutation symmetric upper-polariton state, a localized molecular excitation, and a coherent superposition of a cavity photon and a localized molecular excitation. The extension to a multi-exponential bath correlation decomposition and its numerical illustration are presented in the Supporting Information. The present formulation assumes identical molecular transition energies, identical molecule-cavity couplings within each calculation, identical independent local baths, local diagonal system-bath coupling, one cavity mode, two-level molecules, and the single-excitation manifold. The inhomogeneous static disorder and cavity loss are demonstrated and discussed in the supplementary Information. Within these assumptions, the method provides a direct route to systematically convergent open system dynamics for ensemble sizes that are inaccessible to dense conventional HEOM. Optical response functions~\cite{Mondal2023JCP, Ying2024JCP, Mondal2025JCP1}, higher excitation manifolds~\cite{Mondal2025JCP2, Mondal2026PNAS}, and ensembles of identical multichromophoric units~\cite{Mondal2023JCP} are natural extensions of the same compression strategy.

\section{Model and conventional HEOM}
\label{sec:model}

\subsection{Identical Holstein-Tavis-Cummings Hamiltonian}

\noindent We consider $N$ identical two-level molecules coupled to one cavity mode. We set $\hbar=1$, use the rotating-wave approximation, and restrict the Hilbert space to the single excitation manifold. The ordered basis is
\begin{equation}\label{eq:basis}
\mathcal B_N=\bigl\{|c\rangle,|e_1\rangle,|e_2\rangle,\ldots,|e_N\rangle\bigr\},
\end{equation}
where $|c\rangle$ contains one cavity photon and electronically unexcited molecules, while $|e_i\rangle$ contains an electronic excitation on the $i^{\mathrm{th}}$ molecule  and no cavity photon. The Tavis-Cummings (TC) system Hamiltonian is
\begin{equation}
\hat H_{\mathrm{S}}=\omega_{\mathrm{cav}}|c\rangle\langle c|+\epsilon_{\mathrm x}\sum_{i=1}^{N}|e_i\rangle\langle e_i|+g\sum_{i=1}^{N} \Big(|c\rangle\langle e_i|+|e_i\rangle\langle c|\Big).
\label{eq:HS}
\end{equation}
Here, $\omega_{\mathrm{cav}}$ is the cavity frequency, $\epsilon_{\mathrm x}$ is the molecular transition energy entering the propagated system Hamiltonian, and $g$ is the single-molecule light-matter coupling.  The normalized bright state is
\begin{equation}
|\mathrm{B}\rangle=\frac{1}{\sqrt N}\sum_{i=1}^{N}|e_i\rangle.
\label{eq:bright}
\end{equation}
The collective Rabi splitting on resonance is
\begin{equation}\label{eq:Rabi-splitting}
\Omega_{\mathrm R}=2g\sqrt N.   
\end{equation} 
With $\Delta=\omega_{\mathrm{cav}}-\epsilon_{\mathrm x}$ and $\tan(2\theta)=\Omega_{\mathrm R}/\Delta$, the TC upper and lower polariton eigenstates are,
\begin{subequations}\label{eq:polariton-states}
\begin{align}
|+\rangle &= \sin\theta|\mathrm{B}\rangle + \cos\theta|c\rangle, \label{eq:UP}\\
|-\rangle&= \cos\theta|\mathrm{B}\rangle -\sin\theta|c\rangle,\label{eq:LP}
\end{align}    
with $\theta \in [0, \pi/2]$ being controlled by the cavity detuning. The remaining polaritonic subspace is described by the dark projector
\begin{equation}\label{eq:dark-projector}
\hat{P}_{\mathrm D} = \sum_{i=1}^{N}|e_i\rangle\langle e_i|-|\mathrm{B}\rangle\langle \mathrm{B}|.
\end{equation}
\end{subequations}

\subsection{Local environments and bath decomposition}
\noindent Each molecule is coupled diagonally to an independent harmonic environment. The total Hamiltonian is
\begin{subequations}
\begin{equation}
\hat H=\hat H_{\mathrm{S}}+\hat H_{\mathrm B}+\hat H_{\mathrm{SB}},
\label{eq:total-H}
\end{equation}
with the bath ($\hat{H}_{\mathrm B}$) and the system-bath ($\hat{H}_{\mathrm{SB}}$) interactions defined via
\begin{align}
\hat{H}_{\mathrm B} &= \sum_{i=1}^{N}\sum_{\xi}\omega_{\xi}\hat b_{i\xi}^{\dagger}\hat b_{i\xi},\label{eq:bath-H}
\\
\hat{H}_{\mathrm{SB}} &= \sum_{i=1}^{N}\hat Q_i\otimes\hat B_i,
~~
\hat Q_i=|e_i\rangle\langle e_i|,
\\
\hat B_i&=\sum_{\xi}g_{i\xi}\left(\hat b_{i\xi}^{\dagger}+\hat b_{i\xi}\right).  
\end{align}
\end{subequations} 
For the harmonic environments introduced in Eq.~\ref{eq:bath-H}, the local spectral density on the $i^{\mathrm{th}}$ molecule is
\begin{equation}
J_i(\omega)
=
\pi\sum_{\xi}|g_{i\xi}|^2
\delta(\omega-\omega_{i\xi}).
\label{eq:spectral-density-definition}
\end{equation}
The local baths are mutually independent and statistically identical, \textit{i.e.},
\begin{equation}\label{eq:equal-spectral-density}
J_{i}(\omega)
=
J(\omega)
=
\pi\sum_{\xi}|g_{\xi}|^2
\delta(\omega-\omega_{\xi}).
\end{equation}
The corresponding equilibrium bath correlation functions satisfy
\begin{equation}
\langle \hat B_i(t)\hat B_j(0)\rangle_{\mathrm B}
=
\delta_{ij}C(t).
\label{eq:BCF-independent}
\end{equation}
With the spectral density convention in Eq.~\ref{eq:spectral-density-definition}, the local bath correlation function is
\begin{equation}\label{eq:BCF-spectral}
C(t)
=
\dfrac{1}{\pi}\int_{0}^{\infty}
\mathrm{d}\omega\,
J(\omega)
\left[
\coth\left(\frac{\beta\omega}{2}\right)\cos(\omega t)
-i\sin(\omega t)
\right],
\end{equation}
where $\beta=(k_{\mathrm B}T)^{-1}$. The correlation function entering the finite HEOM is represented as
\begin{subequations}
\label{eq:BCF-exp}
\begin{align}
C_m(t) &= \sum_{k=0}^{m-1} c_k e^{-\nu_k t},
\\
C_m^{*}(t) &= \sum_{k=0}^{m-1} \check{c}_k e^{-\nu_k t}.
\end{align}
\end{subequations}
If the channel $k'$ satisfies $\nu_{k'}=\nu_k^{*}$, then $\check c_k=c_{k'}^{*}$. For a real, self-conjugate rate, $\check c_k=c_k^{*}$. This distinction is required when the correlation function contains conjugate complex decay rates. The coefficients $c_k$ may be complex. Eq.~\ref{eq:BCF-exp} can be obtained from a Matsubara or Pad\'e decomposition or from a controlled numerical fit~\cite{Liu2014JCP, Chen2022JCP, Hu2025JCP, Ying2024JCP}.

\subsection{Conventional scaled HEOM}
\noindent In the conventional HEOM, each auxiliary density operator (ADO) is labeled by a collection of nonnegative integers,
\begin{equation}
\mathbf n=\left(n_{10},\ldots,n_{1,m-1};\ldots;n_{N0},\ldots,n_{N,m-1}\right),
\label{eq:n-label}
\end{equation}
where $n_{ik}\in\mathbb N_0$ is the hierarchy occupation associated with molecule $i\in\{1,\ldots,N\}$ and exponential component $k\in\{0,\ldots,m-1\}$. Its tier and damping rate are
\begin{subequations}
\begin{align}
|\mathbf n|&=\sum_{i=1}^{N}\sum_{k=0}^{m-1}n_{ik},\\
\Gamma_{\mathbf n}&=\sum_{i=1}^{N}\sum_{k=0}^{m-1}n_{ik}\nu_k.
\label{eq:tier-damping}
\end{align}    
\end{subequations}
The physical reduced density operator is the zeroth-tier ADO, $\hat\rho_{\mathrm{S}}(t)=\widetilde{\hat\rho}_{\mathbf 0}(t)$. We use the scaled convention~\cite{Shi2009JCP, Zhu2011JPCB, Bai2024ACR}
\begin{equation}
\widetilde{\hat\rho}_{\mathbf n}=\left[\prod_{i=1}^{N}\prod_{k=0}^{m-1}n_{ik}!\,|c_k|^{n_{ik}}\right]^{-1/2}\hat\rho_{\mathbf n}.
\label{eq:scaled-ADO}
\end{equation}
Let $\mathbf e_{ik}$ denote the hierarchy index that increases $n_{ik}$ by one while leaving all other occupations unchanged. The conventional scaled HEOM is~\cite{Shi2009JCP},
\begin{align}\label{eq:scaled-HEOM}
\frac{\partial}{\partial t}\widetilde{\hat\rho}_{\mathbf n}
={}&-i \Big[\hat H_{\mathrm{S}},\widetilde{\hat\rho}_{\mathbf n} \Big]-\Gamma_{\mathbf n}\widetilde{\hat\rho}_{\mathbf n}\\
&-i\sum_{i=1}^{N}\sum_{k=0}^{m-1}\sqrt{(n_{ik}+1)|c_k|}
\Big[ \hat Q_i,\widetilde{\hat\rho}_{\mathbf n+\mathbf e_{ik}} \Big] \notag\\
&
-i\sum_{i=1}^{N}\sum_{k=0}^{m-1} \sqrt{\frac{n_{ik}}{|c_k|}} \left( c_k\hat Q_i\widetilde{\hat\rho}_{\mathbf n-\mathbf e_{ik}} - \check c_k \widetilde{\hat\rho}_{\mathbf n-\mathbf e_{ik}}\hat Q_i \right).\notag
\end{align}
We retain labels with $|\mathbf{n}|\leq L$. A hard cutoff omits upward connections from tier $L$. There are $Nm$ hierarchy directions, so the number of ADOs through depth $L$ is
\begin{equation}
\mathcal N_{\mathrm{ADO}}(N,m,L) = \binom{Nm+L}{L}.
\label{eq:direct-ado-count}
\end{equation}
Since each ADO is a dense $(N+1)\times(N+1)$ matrix in $\mathcal B_N$, the number of propagated complex variables in the conventional dense representation is
\begin{equation}
\mathcal{N}_{\mathrm{var}}(N,m,L) = \mathcal{N}_{\mathrm{ADO}}(N,m,L)(N+1)^2.
\label{eq:direct-dimension}
\end{equation}

\section{Molecular-relabeling covariance}
\label{sec:covariance}
\noindent Consider a relabeling $\pi$ that assigns the original molecular label $i$ to the new label $\pi(i)$. The corresponding operator satisfies
\begin{subequations}
\begin{align}
\hat U_{\pi}|c\rangle&=|c\rangle,\\
\hat U_{\pi}|e_i\rangle&=|e_{\pi(i)}\rangle.
\label{eq:U-action}
\end{align}
\end{subequations}
Uniform molecular energies and couplings imply
\begin{subequations}
\begin{align}\label{eq:operator-covariance}
\hat U_{\pi}\hat H_{\mathrm{S}}\hat U_{\pi}^{\dagger}&=\hat H_{\mathrm{S}},\\
\hat U_{\pi}\hat Q_i\hat U_{\pi}^{\dagger}&=\hat Q_{\pi(i)}.
\end{align}    
\end{subequations}
Each molecule $i$ is associated with a complete local hierarchy vector
\begin{equation}
\mathbf n_i=(n_{i0},n_{i1},\ldots,n_{i,m-1}),
\end{equation}
where $n_{ik}$ is the occupation number of the $k^{\mathrm{th}}$ exponential term in the bath correlation decomposition for molecule $i$. A molecular relabeling must therefore move the entire vector $\mathbf n_i$, rather than permuting its individual components. If the permutation $\pi$ sends molecule $j$ to the new label $i=\pi(j)$, then the hierarchy vector assigned to position $i$ after relabeling is the vector that previously belonged to molecule $j=\pi^{-1}(i)$. Accordingly, we define
\begin{equation}
(\pi\mathbf n)_{ik}
=
n_{\pi^{-1}(i),k},
\label{eq:hierarchy-relabel}
\end{equation}
for every exponential index $k$. Thus, the relabeling changes only the molecular index and leaves the ordering of the bath exponential components within each local hierarchy vector unchanged. This rearrangement preserves the tier and damping rate,
\begin{equation}
|\pi\mathbf n|=|\mathbf n|,
~~
\Gamma_{\pi\mathbf n}=\Gamma_{\mathbf n},
\label{eq:tier-covariance}
\end{equation}
and maps neighboring labels consistently according to $\pi(\mathbf n\pm\mathbf e_{ik})=\pi\mathbf n\pm\mathbf e_{\pi(i)k}$. Every term in Eq.~\ref{eq:scaled-HEOM} transforms in the same manner. For example, for an arbitrary operator $\hat A$,
\begin{subequations}
\begin{gather}\label{eq:term-covariance}
-i \Big[ \hat H_{\mathrm{S}},\hat U_{\pi}\hat A\hat U_{\pi}^{\dagger} \Big]
=
\hat U_{\pi}\left(-i \Big[\hat H_{\mathrm{S}},\hat A \Big]\right)\hat U_{\pi}^{\dagger},
\\
\Big[ \hat Q_{\pi(i)},\hat U_{\pi}\hat A\hat U_{\pi}^{\dagger} \Big]
=\hat U_{\pi} \Big[\hat Q_i,\hat A \Big]\hat U_{\pi}^{\dagger},
\end{gather}
\begin{align}
c_k\hat Q_{\pi(i)}\hat U_{\pi}\hat A\hat U_{\pi}^{\dagger}
&-\check{c}_k\hat U_{\pi}\hat A\hat U_{\pi}^{\dagger}\hat Q_{\pi(i)}\notag
\\
&=\hat U_{\pi}\left(c_k\hat Q_i\hat A-\check{c}_k\hat A\hat Q_i\right)\hat U_{\pi}^{\dagger}.    
\end{align}   
\end{subequations}
Given any solution $\{\widetilde{\hat\rho}_{\mathbf n}(t)\}$, define the
relabelled hierarchy
\begin{equation}\label{eq:relabelled-hierarchy}
\widetilde{\hat\rho}'_{\pi\mathbf n}(t) = \hat U_{\pi} \widetilde{\hat\rho}_{\mathbf n}(t) \hat U_{\pi}^{\dagger}.
\end{equation}
The identities above show that the primed hierarchy satisfies the same HEOM generator as the original hierarchy, but with correspondingly relabelled initial data. This is molecular-relabeling covariance and generally relates two different initial-value problems.

A relation between ADOs within one propagated hierarchy requires the initial hierarchy itself to satisfy
\begin{equation}
\widetilde{\hat\rho}_{\pi\mathbf n}(0)
=
\hat U_{\pi}
\widetilde{\hat\rho}_{\mathbf n}(0)
\hat U_{\pi}^{\dagger}.
\label{eq:initial-covariance}
\end{equation}
Uniqueness of the finite linear initial-value problem then gives
\begin{equation}
\widetilde{\hat\rho}_{\pi\mathbf n}(t)
=
\hat U_{\pi}
\widetilde{\hat\rho}_{\mathbf n}(t)
\hat U_{\pi}^{\dagger}
\label{eq:central-covariance}
\end{equation}
at every later time.

For a factorized preparation, all higher-tier ADOs initially ($t=0$) vanish. Full compression therefore requires the initial system operator to be invariant under every retained molecular relabeling. If $d$ molecules are distinguished by the initial operator, only relabelings that leave those labels fixed are used, while the remaining $N-d$ molecules are compressed. A general noninvariant initial operator may alternatively be reconstructed from relabelled representative propagations using linearity as discussed later in Sec.~\ref{sec:initial}.

Generator covariance requires identical mean molecular energies, molecule-cavity couplings, local coupling operators, bath decompositions, and a relabeling-compatible hierarchy closure. A generic fixed realization of energetic disorder is not invariant, whereas the ensemble average over independent and identically distributed energetic shifts remains permutation covariant, as shown in the Supporting Information. Molecular transition dipoles do not occur in the HEOM generator written here and thus unequal dipoles may instead break preparation or readout symmetry.



\section{First reduction: Canonical Auxiliary Density Operators}
\label{sec:canonical}
\subsection{One exponential per local bath}

\noindent We begin with the simplest bath decomposition, containing one exponential term per molecule, $m=1$. In this case, each molecule $i$ carries a single nonnegative hierarchy occupation $n_i$, and a molecule-resolved ADO is labeled by
\begin{equation}
\mathbf n=(n_1,n_2,\ldots,n_N).
\end{equation}
The value of $n_i$ specifies the hierarchy occupation associated with the local bath of molecule $i$, while the total hierarchy tier is
\begin{equation}
|\mathbf n|=\sum_{i=1}^{N}n_i.
\end{equation}
Because the molecules and their local baths are identical, simultaneously relabeling the molecules and their hierarchy occupations does not produce an independent dynamical object. Thus, two hierarchy labels that differ only in the ordering of their entries belong to the same relabeling class. For example, the labels $(2,1,0,0)$ and $(0,2,1,0)$ assign the same set of occupations to the molecular ensemble, but attach them to different molecular labels. Their corresponding ADOs are related by the molecular permutation described in Sec.~\ref{sec:covariance} and therefore need not be propagated independently.

We select one representative from each such class by arranging the nonzero occupations in descending order and suppressing the trailing zeros. The resulting canonical occupation pattern is
\begin{equation}
\Lambda=[\Lambda_1,\Lambda_2,\ldots,\Lambda_{r_{\Lambda}}],
~~~
\Lambda_1\geq\Lambda_2\geq\cdots\geq\Lambda_{r_{\Lambda}}>0,
\label{eq:pattern-one}
\end{equation}
where $r_{\Lambda}$ is the number of molecules with nonzero hierarchy occupation. The tier of the pattern is
\begin{equation}
|\Lambda|=\sum_{a=1}^{r_{\Lambda}}\Lambda_a.
\end{equation}
Thus, $\Lambda$ records how the total hierarchy occupation is distributed among the occupied molecular baths, without retaining the physically redundant assignment of those occupations to particular molecular labels. Mathematically, the patterns $\Lambda$ are integer partitions of the hierarchy tier, subject to the condition $r_{\Lambda}\leq N$.

For a system containing $N$ molecules, the canonical pattern is embedded into a molecule resolved hierarchy label by appending the required number of zeros:
\begin{equation}
\mathbf n^{\Lambda}
=
(\Lambda_1,\ldots,\Lambda_{r_{\Lambda}},0,\ldots,0).
\label{eq:n-pattern}
\end{equation}
We retain the ADO associated with this representative and denote it by
\begin{equation}
\hat R_{\Lambda}(t)
=
\widetilde{\hat\rho}_{\mathbf n^{\Lambda}}(t).
\end{equation}
All molecule-resolved ADOs whose hierarchy labels are permutations of $\mathbf n^{\Lambda}$ can be recovered from $\hat R_{\Lambda}(t)$ by applying the corresponding molecular relabeling (Eq.~\ref{eq:U-action}).

For example, the hierarchy labels
\begin{equation}
(2,1,0,0),~~
(0,2,1,0),~~
(1,0,0,2)
~~\longrightarrow~~
\Lambda=[2,1]
\label{eq:pattern-example}
\end{equation}
all reduce to the same canonical pattern. This pattern states that one molecule carries hierarchy occupation $2$, another carries occupation $1$, and all remaining molecules carry occupation $0$. It does not specify which molecular labels carry the two nonzero occupations. The empty pattern $\varnothing$, for which every molecule has zero hierarchy occupation, represents the physical reduced density operator:
\begin{equation}
\hat R_{\varnothing}(t)=\hat\rho_{\mathrm{S}}(t).
\end{equation}
To describe a canonical pattern more conveniently, we introduce the multiplicity $M_q^{\Lambda}$ as the number of molecules carrying hierarchy occupation $q$. For a nonzero occupation $q$, this number is
\begin{equation}
M_q^{\Lambda}
=
\sum_{a=1}^{r_{\Lambda}}
\delta_{\Lambda_a,q},
~~~ q>0,
\label{eq:Mq}
\end{equation}
where $\delta_{\Lambda_a,q}$ is the Kronecker delta. Thus, each occurrence of $q$ in the canonical pattern contributes one to $M_q^{\Lambda}$. Since $r_{\Lambda}$ molecules have nonzero occupations, the remaining $N-r_{\Lambda}$ molecules have zero occupation:
\begin{equation}
M_0^{\Lambda}=N-r_{\Lambda}.
\label{eq:M0}
\end{equation}

The multiplicities satisfy two useful relations. First, every molecule belongs to exactly one occupation category. Second, the sum of all occupations gives the hierarchy tier. Therefore,
\begin{subequations}
\begin{align}
\sum_{q=0}^{L}M_q^{\Lambda} &= N,
\label{eq:multiplicity-molecule-sum}
\\
\sum_{q=1}^{L}qM_q^{\Lambda} &= |\Lambda|,
\label{eq:multiplicity-tier-sum}
\end{align}
\end{subequations}
where $L$ is the maximum retained hierarchy tier.

Although a canonical pattern does not specify which molecule carries each occupation, it represents every molecule-resolved hierarchy label obtained by assigning these occupations to the $N$ molecular labels. The number of distinct assignments is
\begin{equation}
\mathcal N_{\mathrm{ord}}(\Lambda;N)
=
\frac{N!}
{\displaystyle\prod_{q=0}^{L}M_q^{\Lambda}!}.
\label{eq:ordering-count}
\end{equation}
The numerator counts all possible orderings of the $N$ occupations. However, exchanging two molecules that carry the same occupation does not produce a new hierarchy label. The factorial $M_q^{\Lambda}!$ removes these repeated orderings for occupation $q$. This also applies to the zero occupations, so the factor $M_0^{\Lambda}!$ must be included in the denominator.

For example, consider $N=4$ and the canonical pattern $\Lambda=[2,1]$. One molecule carries occupation $2$, one carries occupation $1$, and the remaining two molecules carry occupation $0$. The corresponding multiplicities are
\begin{equation}
M_2^{\Lambda}=1,
~~~~
M_1^{\Lambda}=1,
~~~~
M_0^{\Lambda}=2.
\end{equation}
The number of distinct molecule resolved hierarchy labels represented by this pattern is therefore
\begin{equation}
\mathcal N_{\mathrm{ord}}([2,1];4)
=
\frac{4!}{1!\,1!\,2!}
=
12.
\end{equation}
Thus, instead of independently propagating these twelve relabeling-related ADOs, we retain only the single representative $\hat R_{[2,1]}(t)$.

\subsection{Canonical targets and aligned matrices}

\noindent The equation of motion for an ADO at pattern $\Lambda$ couples it to ADOs in the neighboring hierarchy tiers. These neighboring ADOs are obtained by increasing or decreasing the hierarchy occupation of one molecule. Let $\mathbf e_i$ denote the $N$-component vector whose $i$th entry is one and whose remaining entries are zero. Starting from the representative label $\mathbf n^{\Lambda}$, raising the occupation of molecule $i$ gives
\begin{equation}
\mathbf m_{\Lambda i}^{+}
=
\mathbf n^{\Lambda}+\mathbf e_i,
\label{eq:raw-target-plus}
\end{equation}
whereas lowering its occupation gives
\begin{equation}
\mathbf m_{\Lambda i}^{-}
=
\mathbf n^{\Lambda}-\mathbf e_i,
~~~ n_i^{\Lambda}>0.
\label{eq:raw-target-minus}
\end{equation}
We refer to $\mathbf m_{\Lambda i}^{+}$ and $\mathbf m_{\Lambda i}^{-}$ as raw target labels because their entries remain attached to the original molecular labels.

A raw target label is not necessarily in the descending order used to store canonical representatives. We therefore apply the canonicalization operation $\mathcal C$, which sorts the positive occupations in descending order and removes the trailing zeros. The canonical target patterns are
\begin{subequations}
\begin{align}
\Lambda_i^{+}
&=
\mathcal C\left(\mathbf m_{\Lambda i}^{+}\right),
\label{eq:canonical-target-plus}
\\
\Lambda_i^{-}
&=
\mathcal C\left(\mathbf m_{\Lambda i}^{-}\right).
\label{eq:canonical-target-minus}
\end{align}
\end{subequations}
Only the representatives $\hat R_{\Lambda_i^{+}}$ and $\hat R_{\Lambda_i^{-}}$ associated with these canonical patterns are stored and propagated.

Canonicalization may change the assignment of occupations to molecular labels. This distinction matters because the corresponding HEOM term contains the local system operator $\hat Q_i$ of the molecule whose occupation was raised or lowered. Before using a stored target ADO in that term, its molecular indices must therefore be returned to the ordering of the raw target label.

For example, consider $N=4$ and $\Lambda=[2,2]$, with representative label
\begin{equation}
\mathbf n^{\Lambda}=(2,2,0,0).
\end{equation}
Lowering the occupation of molecule $1$ produces the raw target
\begin{equation}
\mathbf m_{\Lambda 1}^{-}=(1,2,0,0).
\end{equation}
After sorting, the corresponding canonical target is
\begin{equation}
\Lambda_1^{-}=[2,1],
~~~
\mathbf n^{\Lambda_1^{-}}=(2,1,0,0).
\end{equation}
The stored target places occupation $2$ on molecule $1$ and occupation $1$ on molecule $2$, whereas the raw target requires the opposite assignment. Consequently, molecules $1$ and $2$ must be exchanged before this stored representative can be used in the lowering term associated with molecule $1$.

More generally, let $\pi_{\Lambda i}^{\pm}$ denote a molecular relabeling that converts the stored canonical ordering of $\Lambda_i^{\pm}$ into the molecular ordering of the raw target $\mathbf m_{\Lambda i}^{\pm}$. The target ADO expressed in the required raw ordering is then
\begin{equation}
\hat R_{\Lambda_i^{\pm}}^{[\mathbf m_{\Lambda i}^{\pm}]}
=
\hat U_{\pi_{\Lambda i}^{\pm}}
\hat R_{\Lambda_i^{\pm}}
\hat U_{\pi_{\Lambda i}^{\pm}}^{\dagger}.
\label{eq:aligned-target}
\end{equation}
We call this the aligned target matrix. It is obtained by relabeling a stored representative and is not a separate ADO that must be propagated.

Using these aligned targets, the equation of motion can be written entirely in terms of canonical representatives. For a hierarchy truncated at tier $L$, we define
\begin{equation}
\chi_L(\Lambda)
=
\begin{cases}
1, & |\Lambda|<L,\\
0, & |\Lambda|=L.
\end{cases}
\label{eq:hard-cutoff-factor}
\end{equation}
This factor prevents raising operations from generating ADOs above the maximum retained tier. The one-exponential canonical-ADO equation is
\begin{align}\label{eq:canonical-ADO-eom}
\frac{\partial}{\partial t}\hat R_{\Lambda}
={}&
-i \Big[ \hat H_{\mathrm{S}},\hat R_{\Lambda} \Big]
-\nu|\Lambda|\hat R_{\Lambda}
\\
&-i\chi_L(\Lambda)
\sum_{i=1}^{N}
\sqrt{(n_i^{\Lambda}+1)|c|}
\left[
\hat Q_i,
\hat R_{\Lambda_i^{+}}^{[\mathbf m_{\Lambda i}^{+}]}
\right]
\notag\\
&-i
\sum_{\substack{i=1\\n_i^{\Lambda}>0}}^{N}
\sqrt{\frac{n_i^{\Lambda}}{|c|}}
\left(
c\hat Q_i
\hat R_{\Lambda_i^{-}}^{[\mathbf m_{\Lambda i}^{-}]}
-
c^{*}
\hat R_{\Lambda_i^{-}}^{[\mathbf m_{\Lambda i}^{-}]}
\hat Q_i
\right).\notag
\end{align}
The first line contains the system evolution and the hierarchy damping. The second line couples $\hat R_{\Lambda}$ to the next higher tier by raising one molecular occupation. The final two lines couple it to the next lower tier by lowering a nonzero occupation. In every coupling term, the target representative is first aligned with the molecular ordering required by the corresponding local operator $\hat Q_i$.

Eq.~\ref{eq:canonical-ADO-eom} is obtained directly from the conventional HEOM using the permutation covariance relation in Eq.~\ref{eq:central-covariance}. Canonicalization and alignment only identify and reconstruct relabeling related ADOs. They introduce no additional approximation relative to the conventional HEOM with the same bath decomposition, hierarchy depth, and closure.

We next count the number of ADOs retained after this first symmetry reduction. At tier $\ell$, every canonical pattern is a descending list of positive integers whose sum is $\ell$ and whose length does not exceed $N$. The number of such patterns is the restricted integer partition number $p_N(\ell)$. Including all tiers from zero through $L$, the total number of canonical representatives is
\begin{equation}
\mathcal N_{\mathrm{ADO}}^{\mathrm{can}}(N,m=1,L)
=
\sum_{\ell=0}^{L}p_N(\ell).
\label{eq:canonical-count-one}
\end{equation}
A pattern at tier $\ell$ can contain at most $\ell$ nonzero entries because every nonzero occupation is at least one. Since $\ell\leq L$, no retained pattern can involve more than $L$ molecules with nonzero occupations. Therefore, once $N\geq L$, increasing the number of molecules does not generate any additional canonical patterns. In this regime, the restriction on the pattern length becomes irrelevant, and
\begin{equation}\label{eq:first-saturation}
\mathcal N_{\mathrm{ADO}}^{\mathrm{can}}(N,m=1,L)
=
\sum_{\ell=0}^{L}p(\ell),
~~~~ N\geq L,
\end{equation}
where $p(\ell)$ is the unrestricted integer partition number.
\begin{figure}[!h]
    \centering
    \includegraphics[width=0.85\linewidth]{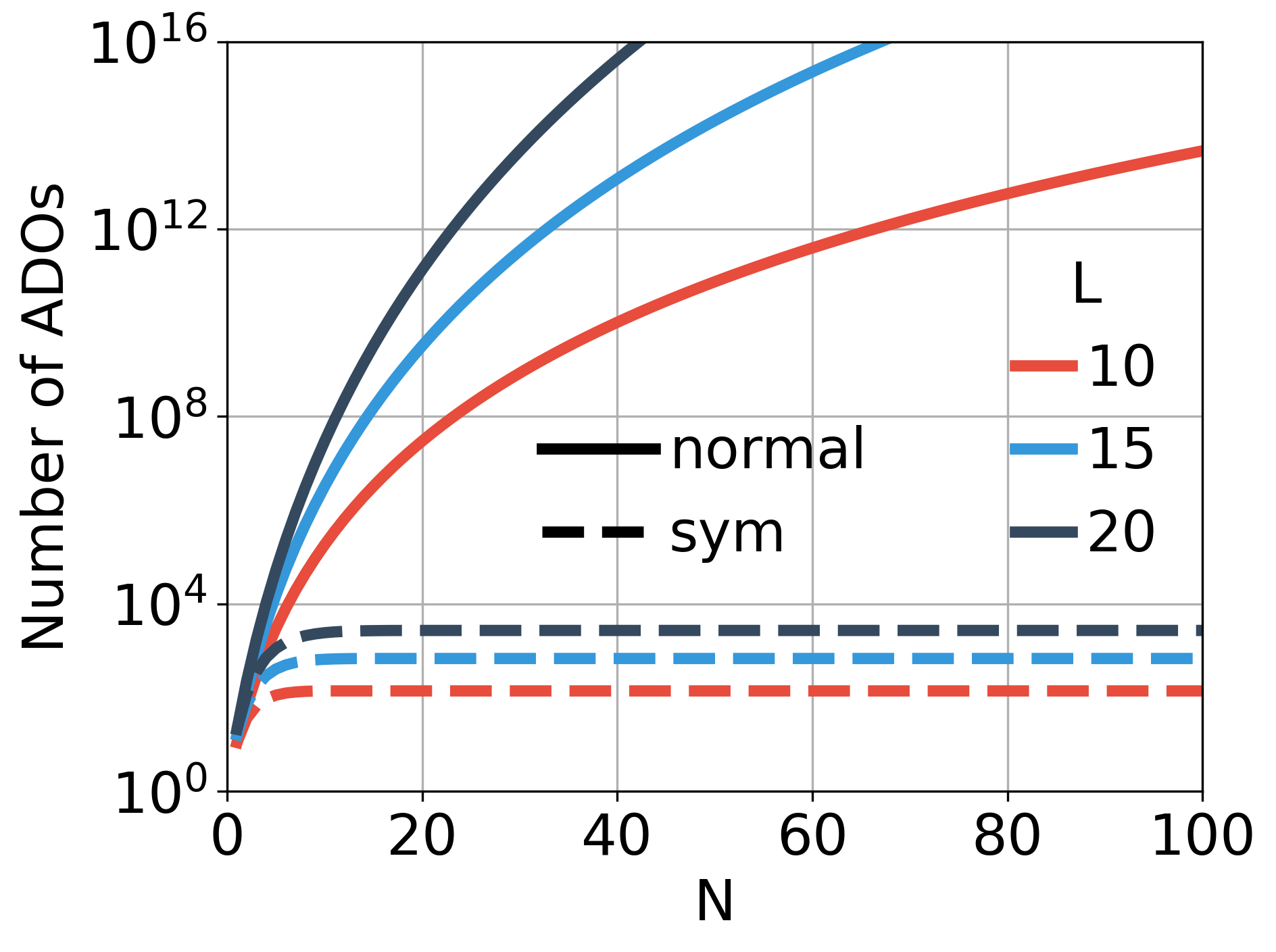}
    \caption{Number of ADOs retained through hierarchy depth $L$ for a one-exponential decomposition per local bath. The curves labeled ``normal'' correspond to the molecule-resolved hierarchy used by conventional HEOM, whereas those labeled ``sym'' show the canonical ADO representatives obtained after molecular relabeling. The conventional hierarchy grows rapidly with the ensemble size $N$, while the number of canonical representatives saturates once $N\geq L$.}
    \label{fig:ADO_count_normal_vs_sym}
\end{figure}

Fig.~\ref{fig:ADO_count_normal_vs_sym} illustrates the first source of computational savings. The molecule resolved hierarchy of conventional HEOM grows combinatorially with $N$, whereas canonicalization removes hierarchy labels that differ only by relabeling identical molecules. At fixed $L$, no retained pattern can contain more than $L$ nonzero molecular occupations, so the number of canonical representatives becomes independent of $N$ for $N\geq L$. This saturation applies only to the number of ADO representatives. If every canonical ADO is stored as a full $(N+1)\times(N+1)$ matrix, its memory and propagation costs continue to increase quadratically with $N$. The unique-variable reduction introduced below removes this remaining matrix redundancy.

\section{Second reduction: unique variables and matrix-free propagation}
\label{sec:variables}

\noindent The first symmetry reduction replaces all relabeling-related ADOs by a single canonical representative $\hat R_{\Lambda}$. However, each representative is still an $(N+1)\times(N+1)$ matrix in the single excitation basis. Storing the full matrix would therefore leave a computational cost that grows quadratically with the number of molecules.

A second reduction is possible because many matrix elements within a canonical representative are identical. We first identify these repeated entries and count the number of unique variables. We then derive equations that propagate the unique variables directly, without reconstructing the full ADO matrices.

\subsection{Repeated matrix elements within a representative}

\noindent Consider two molecules $i$ and $j$ that carry the same hierarchy occupation in the representative label $\mathbf n^{\Lambda}$:
\begin{equation}
n_i^{\Lambda}=n_j^{\Lambda}.
\end{equation}
Exchanging these two molecules leaves the representative hierarchy label unchanged. The permutation covariance relation in Eq.~\ref{eq:central-covariance} therefore gives
\begin{equation}
\hat R_{\Lambda}
=
\hat U_{ij}
\hat R_{\Lambda}
\hat U_{ij}^{\dagger},
~~~
n_i^{\Lambda}=n_j^{\Lambda}.
\label{eq:within-representative}
\end{equation}
Consequently, simultaneously exchanging the corresponding molecular rows and columns cannot change the ADO matrix.

This means that a matrix element does not depend on the individual labels of the molecules involved. It depends only on the hierarchy occupation categories carried by its row and column. Molecules with the same occupation are therefore indistinguishable within the representative ADO.

This symmetry does not imply that $\hat R_{\Lambda}$ is Hermitian. Although the physical reduced density operator is Hermitian, higher-tier ADOs are generally non-Hermitian. The matrix elements associated with the two opposite directions of a coherence must therefore be retained as independent complex variables.

Let $\mathcal G_{\Lambda}$ denote the list of occupation categories present in $\mathbf n^{\Lambda}$, including the zero-occupation category, and let
\begin{equation}
G_{\Lambda}=|\mathcal G_{\Lambda}|
\end{equation}
be the number of such categories. For each $q\in\mathcal G_{\Lambda}$, we define the cavity population variable
\begin{equation}
Z^{\Lambda}
=
\langle c|
\hat R_{\Lambda}
|c\rangle,
\label{eq:Z-variable}
\end{equation}
the two cavity-molecule coherence variables
\begin{subequations}
\begin{align}
X_q^{\Lambda}
&=
\langle e_i|
\hat R_{\Lambda}
|c\rangle,
~~~ n_i^{\Lambda}=q,
\label{eq:X-variable}
\\
Y_q^{\Lambda}
&=
\langle c|
\hat R_{\Lambda}
|e_i\rangle,
~~~ n_i^{\Lambda}=q,
\label{eq:Y-variable}
\end{align}
\end{subequations}
and the molecular population variable
\begin{equation}
P_q^{\Lambda}
=
\langle e_i|
\hat R_{\Lambda}
|e_i\rangle,
~~~ n_i^{\Lambda}=q.
\label{eq:P-variable}
\end{equation}
Because every molecule in category $q$ is equivalent within the representative, the definitions do not depend on which particular molecule $i$ is selected from that category.

For two different occupation categories $q$ and $r$, we define the ordered intermolecular coherence
\begin{equation}
S_{q,r}^{\Lambda}
=
\langle e_i|
\hat R_{\Lambda}
|e_j\rangle,
~~~
n_i^{\Lambda}=q,
~~~
n_j^{\Lambda}=r,
~~~
q\neq r.
\label{eq:S-variable}
\end{equation}
The ordering of the category indices matters because higher-tier ADOs need not be Hermitian. In general,
\begin{equation}
S_{q,r}^{\Lambda}
\neq
S_{r,q}^{\Lambda}.
\end{equation}
If category $q$ contains at least two molecules, we must also retain the off-diagonal coherence between two different molecules in that same category:
\begin{equation}
T_q^{\Lambda}
=
\langle e_i|
\hat R_{\Lambda}
|e_j\rangle,
~~~
i\neq j,
~~~
n_i^{\Lambda}=n_j^{\Lambda}=q.
\label{eq:T-variable}
\end{equation}
The variable $T_q^{\Lambda}$ exists only when the multiplicity satisfies
\begin{equation}
M_q^{\Lambda}\geq2.
\end{equation}
As an explicit example, consider $N=4$ and the canonical pattern $\Lambda=[2,1]$. Its representative molecule-resolved label is
\begin{equation}
\mathbf n^{[2,1]}=(2,1,0,0).
\end{equation}
The three occupation categories are therefore $2$, $1$, and $0$. In the ordered basis
\begin{equation}
\mathcal B_4
=
\{
|c\rangle,
|e_1\rangle,
|e_2\rangle,
|e_3\rangle,
|e_4\rangle
\},
\end{equation}
the representative has the structure
\begin{equation}
\hat R_{[2,1]}
\longrightarrow
\begin{pmatrix}
Z&Y_2&Y_1&Y_0&Y_0\\
X_2&P_2&S_{2,1}&S_{2,0}&S_{2,0}\\
X_1&S_{1,2}&P_1&S_{1,0}&S_{1,0}\\
X_0&S_{0,2}&S_{0,1}&P_0&T_0\\
X_0&S_{0,2}&S_{0,1}&T_0&P_0
\end{pmatrix}_{\mathcal B_4}^{[2,1]}.
\label{eq:matrix-example}
\end{equation}
The superscript $[2,1]$ is understood to apply to every variable on the right-hand side. The full matrix contains $25$ entries, but only $17$ of them are distinct. For example, molecules $3$ and $4$ both belong to the zero-occupation category. Their cavity coherences are therefore both described by $X_0$ and $Y_0$, their populations are both described by $P_0$, and their mutual coherences are described by $T_0$.

The advantage becomes more pronounced as $N$ increases. For the same pattern $\Lambda=[2,1]$ with $N=10$, the full matrix contains
\begin{equation}
(N+1)^2=11^2=121
\end{equation}
entries. Nevertheless, the occupation categories remain $2$, $1$, and $0$, so the number of unique variables remains $17$.

\subsection{Number of unique variables and saturation}

\noindent We now count the unique variables associated with a general canonical pattern. Let $h_{\Lambda}$ be the number of occupation categories containing at least two molecules:
\begin{equation}
h_{\Lambda}
=
\sum_{q\in\mathcal G_{\Lambda}}
\Theta\left(M_q^{\Lambda}-2\right),
\label{eq:h-Lambda}
\end{equation}
where $\Theta(x)$ equals one for $x\geq0$ and zero otherwise. Each such category contributes one variable of type $T_q^{\Lambda}$. For a pattern containing $G_{\Lambda}$ occupation categories, the unique variables consist of:
\begin{enumerate}
\item one cavity population variable $Z^{\Lambda}$;
\item $G_{\Lambda}$ variables of each type $X_q^{\Lambda}$, $Y_q^{\Lambda}$, and $P_q^{\Lambda}$;
\item $G_{\Lambda}(G_{\Lambda}-1)$ ordered variables $S_{q,r}^{\Lambda}$ with $q\neq r$;
\item $h_{\Lambda}$ same-category coherences $T_q^{\Lambda}$.
\end{enumerate}
The number of distinct complex variables in $\hat R_{\Lambda}$ is therefore
\begin{align}
u_{\mathrm{var}}(\Lambda;N,1)
&=
1
+
3G_{\Lambda}
+
G_{\Lambda}(G_{\Lambda}-1)
+
h_{\Lambda}
\notag\\
&=
(G_{\Lambda}+1)^2+h_{\Lambda}.
\label{eq:uvar-count}
\end{align}
Summing over all canonical patterns through hierarchy depth $L$ gives the total number of unique variables
\begin{equation}
\mathcal{N}_{\mathrm{var}}^{\mathrm{unq}}(N,1,L)
=
\sum_{|\Lambda|\leq L}
\left[
(G_{\Lambda}+1)^2+h_{\Lambda}
\right].
\label{eq:Du-one}
\end{equation}
\begin{figure}[!h]
    \centering
    \includegraphics[width=0.85\linewidth]{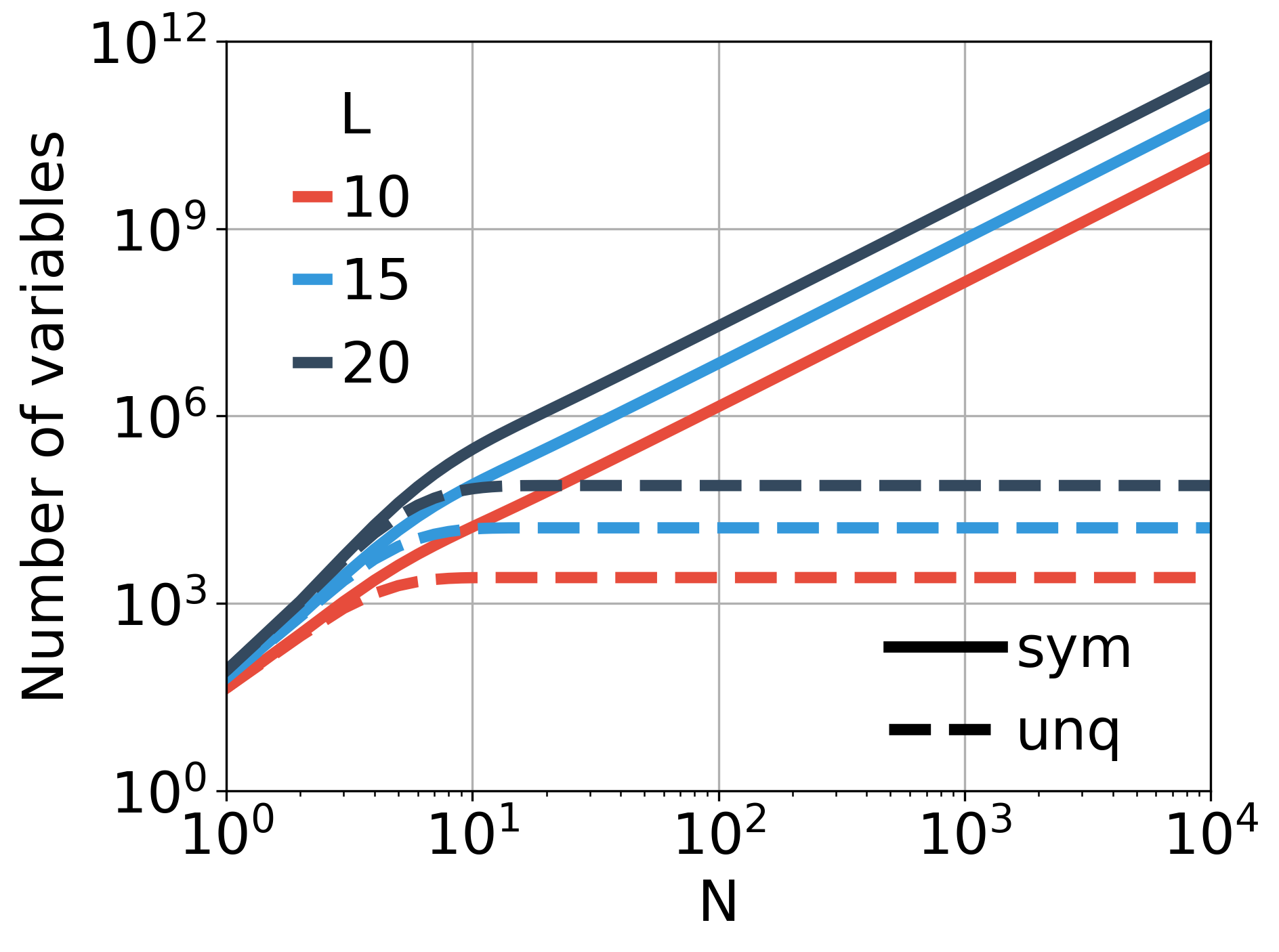}
    \caption{Number of propagated complex variables through hierarchy depth $L$ for a one-exponential decomposition per local bath. Curves labeled ``sym'' correspond to storing every entry of the full $(N+1)\times(N+1)$ matrix associated with each canonical ADO; curves labeled ``unq'' show the dimension after retaining only the unique variables. The full-matrix dimension continues to grow with $N$, whereas the unique-variable dimension saturates for $N\geq L+2$.}
    \label{fig:uvar_count_sym_vs_unq}
\end{figure}

Fig.~\ref{fig:uvar_count_sym_vs_unq} demonstrates the additional reduction obtained by propagating unique variables. Although the canonical-ADO count has already saturated for $N\geq L$, retaining full ADO matrices preserves an $(N+1)^2$ dependence. The unique variable representation instead depends only on the occupation categories present in each canonical pattern and on whether a category contains one or at least two molecules. For a fixed pattern, the dependence on $N$ enters only through the number of molecules in the zero-occupation category,
\begin{equation}
M_0^{\Lambda}=N-r_{\Lambda}.
\end{equation}
Once $M_0^{\Lambda}\geq2$, adding another zero-occupation molecule changes only the value of this multiplicity. It does not introduce a new occupation category or a new type of unique variable. Every pattern retained through depth $L$ satisfies
\begin{equation}
r_{\Lambda}\leq L.
\end{equation}
Therefore, if $N\geq L+2$, even a pattern containing the maximum possible number $L$ of nonzero occupations has at least two molecules in the zero category. All canonical patterns have then reached their final list of unique variable types. Consequently,
\begin{equation}
\mathcal{N}_{\mathrm{var}}^{\mathrm{unq}}(N,1,L)
=
\mathcal{N}_{\mathrm{var}}^{\mathrm{unq}}(L+2,1,L),
~~~
N\geq L+2.
\label{eq:second-saturation}
\end{equation}
The first symmetry reduction makes the number of canonical ADO representatives independent of $N$ for $N\geq L$. The second reduction goes further and it makes the total number of propagated complex variables independent of $N$ for $N\geq L+2$. The saturated dimension can still grow rapidly with the hierarchy depth $L$ and with the number of exponential terms in the bath correlation decomposition.

\subsection{Direct propagation of the unique variables}
\label{sec:direct-uvar}

\noindent The unique variable representation becomes computationally useful only if its variables can be propagated without reconstructing the full ADO matrices. We therefore rewrite the HEOM directly in terms of $Z$, $X$, $Y$, $P$, $S$, and $T$.

There are two types of contributions to the derivative of a unique variable. The system Hamiltonian couples matrix elements within the same canonical representative. The hierarchy terms connect the representative to canonical patterns in the neighboring hierarchy tiers. We consider these two contributions separately.

\subsubsection{Global vector of propagated variables}

\noindent For every canonical pattern $\Lambda$, we collect its unique variables into a local vector $|R_{\Lambda}(t)\rangle$. This is a bookkeeping vector rather than a quantum state. Its entries are the unique variables belonging to $\Lambda$, placed in a fixed order. The complete propagated vector is obtained by stacking these local vectors according to a fixed ordering of the canonical patterns:
\begin{equation}
|R(t)\rangle
=
\begin{pmatrix}
|R_{\varnothing}(t)\rangle \\[4pt]
|R_{[1]}(t)\rangle \\[4pt]
|R_{[2]}(t)\rangle \\[4pt]
|R_{[1,1]}(t)\rangle \\[4pt]
\vdots
\end{pmatrix}.
\label{eq:global-R}
\end{equation}
Thus, every entry of $|R(t)\rangle$ is identified by its canonical pattern and its variable type within that pattern. Because the HEOM are linear, the evolution can be written formally as
\begin{equation}
\frac{\partial}{\partial t}|R(t)\rangle
=
\hat{\mathcal M}_{u}|R(t)\rangle.
\label{eq:uvar-propagator}
\end{equation}
The generator $\hat{\mathcal M}_{u}$ is not constructed as a dense matrix. Instead, its action is evaluated directly from precomputed connections between the unique variables. The following derivation explains how these connections are obtained.

\subsubsection{Hamiltonian terms from grouped molecular sums}

\noindent The cavity-molecule coupling in $\hat H_{\mathrm{S}}$ generates sums over molecular rows and columns. In a full matrix calculation, these sums would contain one term for every molecule. In the unique variable representation, all molecules in the same occupation category contribute identical entries. Their contributions can therefore be combined using the multiplicities $M_q^{\Lambda}$.

Consider any molecule $i$ belonging to category $q$. The sum over the molecular columns of its row is
\begin{equation}
\mathcal R_q^{\Lambda}
=
\sum_{j=1}^{N}
\langle e_i|
\hat R_{\Lambda}
|e_j\rangle,
~~~ n_i^{\Lambda}=q.
\label{eq:row-sum-definition}
\end{equation}
This sum contains the diagonal entry $P_q^{\Lambda}$, the off-diagonal entries connecting molecule $i$ to other molecules in category $q$, and the entries connecting it to molecules in every different category. Grouping identical terms gives
\begin{equation}
\mathcal R_q^{\Lambda}
=
P_q^{\Lambda}
+
\left(M_q^{\Lambda}-1\right)T_q^{\Lambda}
+
\sum_{\substack{r\in\mathcal G_{\Lambda}\\r\neq q}}
M_r^{\Lambda}S_{q,r}^{\Lambda}.
\label{eq:grouped-row-sum}
\end{equation}
Similarly, the sum over the molecular rows of column $i$ is
\begin{equation}
\mathcal C_q^{\Lambda}
=
\sum_{j=1}^{N}
\langle e_j|
\hat R_{\Lambda}
|e_i\rangle,
~~~ n_i^{\Lambda}=q,
\label{eq:column-sum-definition}
\end{equation}
which becomes
\begin{equation}
\mathcal C_q^{\Lambda}
=
P_q^{\Lambda}
+
\left(M_q^{\Lambda}-1\right)T_q^{\Lambda}
+
\sum_{\substack{r\in\mathcal G_{\Lambda}\\r\neq q}}
M_r^{\Lambda}S_{r,q}^{\Lambda}.
\label{eq:grouped-column-sum}
\end{equation}
If $M_q^{\Lambda}=1$, no second molecule exists in category $q$, and the term containing $T_q^{\Lambda}$ is absent. For the example $\Lambda=[2,1]$ with $N=4$, the zero-category contains two molecules. The row sum associated with a zero-occupation molecule is
\begin{equation}
\mathcal R_0^{[2,1]}
=
P_0^{[2,1]}
+
T_0^{[2,1]}
+
S_{0,2}^{[2,1]}
+
S_{0,1}^{[2,1]}.
\label{eq:row-sum-example}
\end{equation}
The term $T_0^{[2,1]}$ accounts for the other molecule in the zero-category. The two $S$ variables account for the molecules in categories $2$ and $1$.

For a larger system, many identical terms are absorbed into the multiplicity factors. For example, if category $r$ contains one hundred molecules, its entire contribution to $\mathcal R_q^{\Lambda}$ is evaluated as $100S_{q,r}^{\Lambda}$ rather than as a sum over one hundred separately stored entries.

For the one-exponential hierarchy, the damping rate of pattern $\Lambda$ is
\begin{equation}
\Gamma_{\Lambda}
=
\nu|\Lambda|.
\label{eq:Gamma-Lambda-one}
\end{equation}
Using the grouped sums, the Hamiltonian and damping contributions to two representative variables are
\begin{subequations}
\begin{align}
\left.
\frac{\partial Z^{\Lambda}}{\partial t}
\right|_{H+\Gamma}
={}&
-ig
\sum_{q\in\mathcal G_{\Lambda}}
M_q^{\Lambda}
\left(
X_q^{\Lambda}-Y_q^{\Lambda}
\right)
-\Gamma_{\Lambda}Z^{\Lambda},
\label{eq:HZ}
\\
\left.
\frac{\partial X_q^{\Lambda}}{\partial t}
\right|_{H+\Gamma}
={}&
-i\left(
\epsilon_{\mathrm x}-\omega_{\mathrm{cav}}
\right)X_q^{\Lambda}
-ig\left(
Z^{\Lambda}-\mathcal R_q^{\Lambda}
\right)
-\Gamma_{\Lambda}X_q^{\Lambda}.
\label{eq:HX}
\end{align}
\end{subequations}
The first equation combines the identical contributions of the $M_q^{\Lambda}$ molecules belonging to category $q$. The second replaces a sum over all molecular columns by the grouped row sum $\mathcal R_q^{\Lambda}$. The remaining component equations are obtained in the same manner and are given in the Supporting Information.

\subsubsection{Hierarchy terms and aligned target access}

\noindent The upward and downward hierarchy terms connect variables belonging to different canonical patterns. Changing the occupation of molecule $i$ produces the raw targets $\mathbf m_{\Lambda i}^{+}$ and $\mathbf m_{\Lambda i}^{-}$ introduced in Eqs.~\ref{eq:raw-target-plus} and \eqref{eq:raw-target-minus}. Canonicalizing these raw labels identifies the stored target patterns $\Lambda_i^{+}$ and $\Lambda_i^{-}$.

The ordering of the molecules in a stored target representative may differ from the ordering required by the raw hierarchy term. Formally, the required target matrix is obtained by aligning the stored representative:
\begin{equation}
\hat R_{\Lambda'}^{[\mathbf m]}
=
\hat U_{\pi}
\hat R_{\Lambda'}
\hat U_{\pi}^{\dagger},
~~~
\Lambda'=\mathcal C(\mathbf m).
\label{eq:generic-aligned-target}
\end{equation}
Explicitly constructing this matrix would restore the $(N+1)^2$ storage that the unique variable reduction is intended to remove. Instead, we access only the individual matrix entry required by a particular component equation.

For a raw target $\mathbf m$ with canonical pattern $\Lambda'$, define the aligned target accessor
\begin{equation}
\mathcal A_{\Lambda'}^{[\mathbf m]}(a,b)
=
\langle a|
\hat R_{\Lambda'}^{[\mathbf m]}
|b\rangle,
~~~
|a\rangle,|b\rangle\in\mathcal B_N.
\label{eq:aligned-accessor}
\end{equation}
The accessor first maps the row and column labels into the canonical ordering of $\Lambda'$. It then determines whether the requested entry is represented by a $Z$, $X$, $Y$, $P$, $S$, or $T$ variable and reads that variable directly from the propagated vector. It does not construct the full target matrix.

For targets generated from pattern $\Lambda$ by changing the occupation of molecule $i$, we use the shorter notation
\begin{equation}
\mathcal A_{\Lambda i}^{\pm}(a,b)
\equiv
\mathcal A_{\Lambda_i^{\pm}}^{[\mathbf m_{\Lambda i}^{\pm}]}(a,b).
\label{eq:short-accessor}
\end{equation}
To derive the hierarchy contribution to a general unique variable, write
\begin{equation}
V_{\alpha}^{\Lambda}
=
\langle a_{\alpha}^{\Lambda}|
\hat R_{\Lambda}
|b_{\alpha}^{\Lambda}\rangle.
\label{eq:generic-uvar}
\end{equation}
The index $\alpha$ identifies one of the retained $Z$, $X$, $Y$, $P$, $S$, or $T$ variables, while $a_{\alpha}^{\Lambda}$ and $b_{\alpha}^{\Lambda}$ specify its row and column basis states.

The local system-bath coupling operator is $\hat Q_i
=
|e_i\rangle\langle e_i|$. For any matrix $\hat A$, its commutator with $\hat Q_i$ satisfies
\begin{equation}
\langle a|
\big[\hat Q_i,\hat A\big]
|b\rangle
=
\delta_{a,e_i}
\langle e_i|\hat A|b\rangle
-
\delta_{b,e_i}
\langle a|\hat A|e_i\rangle.
\label{eq:projector-commutator-element}
\end{equation}
This identity has a simple interpretation. The first term is present only when the row of the propagated variable corresponds to molecule $i$. The second term is present only when its column corresponds to molecule $i$. Using Eq.~\ref{eq:projector-commutator-element}, the upward hierarchy contribution is
\begin{align}\label{eq:uvar-up}
\left.
\frac{\partial V_{\alpha}^{\Lambda}}{\partial t}
\right|_{\mathrm{up}}
={}&
-i\chi_L(\Lambda)
\sum_{i=1}^{N}
\sqrt{
\left(n_i^{\Lambda}+1\right)|c|
}
\\
&\times
\Big[
\delta_{a_{\alpha}^{\Lambda},e_i}
\mathcal A_{\Lambda i}^{+}
\left(
e_i,b_{\alpha}^{\Lambda}
\right)
-
\delta_{b_{\alpha}^{\Lambda},e_i}
\mathcal A_{\Lambda i}^{+}
\left(
a_{\alpha}^{\Lambda},e_i
\right)
\Big].\notag
\end{align}
The factor $\chi_L(\Lambda)$ prevents an upward connection when $|\Lambda|=L$. For the downward contribution, we define
\begin{equation}
w_i^{\Lambda}
=
\sqrt{\frac{n_i^{\Lambda}}{|c|}}.
\label{eq:w-one-exp}
\end{equation}
Only molecules with nonzero hierarchy occupation can be lowered. The downward contribution is
\begin{align}
\left.
\frac{\partial V_{\alpha}^{\Lambda}}{\partial t}
\right|_{\mathrm{down}}
={}&
-i
\sum_{\substack{i=1\\n_i^{\Lambda}>0}}^{N}
w_i^{\Lambda}c\,
\delta_{a_{\alpha}^{\Lambda},e_i}
\mathcal A_{\Lambda i}^{-}
\left(
e_i,b_{\alpha}^{\Lambda}
\right)
\notag\\
&+
i
\sum_{\substack{i=1\\n_i^{\Lambda}>0}}^{N}
w_i^{\Lambda}c^{*}\,
\delta_{b_{\alpha}^{\Lambda},e_i}
\mathcal A_{\Lambda i}^{-}
\left(
a_{\alpha}^{\Lambda},e_i
\right).
\label{eq:uvar-down}
\end{align}
The apparently large molecular sums in Eqs.~\ref{eq:uvar-up} and \eqref{eq:uvar-down} contain very few nonzero terms. Their action can be understood by considering the type of the propagated variable:

\begin{itemize}
\item For $Z^{\Lambda}=\langle c|\hat R_{\Lambda}|c\rangle$, neither the row nor the column is a molecular excitation. Both Kronecker deltas vanish, so there is no direct hierarchy contribution.
\item For $X_q^{\Lambda}=\langle e_i|\hat R_{\Lambda}|c\rangle$, only the row delta is nonzero. The hierarchy term therefore requires one aligned target entry.
\item For $Y_q^{\Lambda}=\langle c|\hat R_{\Lambda}|e_i\rangle$, only the column delta is nonzero. Again, only one aligned target entry is required.
\item For $P_q^{\Lambda}$, $S_{q,r}^{\Lambda}$, or $T_q^{\Lambda}$, both basis labels are molecular excitations. At most one row contribution and one column contribution can survive.
\end{itemize}
Thus, every unique variable requires at most two aligned target accesses in each hierarchy direction. The sums over $N$ molecules are useful for writing the equations compactly, but they are not evaluated as molecule-sized loops during time propagation.

\subsubsection{Precomputed matrix-free right-hand side}

\noindent Before propagation begins, every nonzero contribution to the derivative of every unique variable is identified. For each contribution, the precomputation records the source variable, the target variable, the alignment required to access it, its multiplicity, and its numerical coefficient.

The right-hand side evaluation then requires only three operations: read the appropriate entries of the global vector, multiply them by their precomputed coefficients, and add the results to the derivatives. The full ADO matrices, molecular permutation matrices, and dense generator $\hat{\mathcal M}_{u}$ are never constructed.

After the saturation condition $N\geq L+2$ is reached, increasing $N$ changes only multiplicity factors such as $M_q^{\Lambda}$. It does not enlarge the canonical-pattern list, the unique variable vector, or the number of runtime connections. The storage requirement and the right-hand-side (RHS) operation count therefore become independent of the molecular ensemble size.

\subsection{Physical observables}

\noindent The physical reduced density operator is the zeroth-tier representative,
\begin{equation}
\hat\rho_{\mathrm{S}}(t)
=
\hat R_{\varnothing}(t).
\end{equation}
For $N\geq2$, the empty pattern contains only the zero-occupation category, with
\begin{equation}
M_0^{\varnothing}=N.
\end{equation}
The complete physical density operator is therefore specified by only five unique variables:
\begin{equation}
Z^{\varnothing},
~~~
X_0^{\varnothing},
~~~
Y_0^{\varnothing},
~~~
P_0^{\varnothing},
~~~
T_0^{\varnothing}.
\end{equation}
The trace, cavity population, and total molecular-excitation population are
\begin{subequations}
\begin{align}
\mathrm{Tr}\left[\hat\rho_{\mathrm{S}}\right]
&=
Z^{\varnothing}
+
NP_0^{\varnothing},
\label{eq:trace-uvar}
\\
P_{\mathrm{cav}}
&=
\langle c|
\hat\rho_{\mathrm{S}}
|c\rangle
=
Z^{\varnothing},
\label{eq:Pcav-uvar}
\\
P_{\mathrm{exc}}
&=
\sum_{i=1}^{N}
\langle e_i|
\hat\rho_{\mathrm{S}}
|e_i\rangle
=
NP_0^{\varnothing}.
\label{eq:Pexc-uvar}
\end{align}
The bright state population is
\begin{equation}
P_{\mathrm B}
=
\langle \mathrm B|
\hat\rho_{\mathrm{S}}
|\mathrm B\rangle
=
P_0^{\varnothing}
+
(N-1)T_0^{\varnothing}.
\label{eq:PB-uvar}
\end{equation}
The total population in the dark molecular subspace is
\begin{equation}
P_{\mathrm D}
=
\mathrm{Tr}\left[
\hat\rho_{\mathrm{S}}\hat P_{\mathrm D}
\right]
=
(N-1)
\left(
P_0^{\varnothing}-T_0^{\varnothing}
\right).
\label{eq:PD-uvar}
\end{equation}
Using the mixing convention in Eq.~\ref{eq:polariton-states}, the upper- and lower-polariton populations are
\begin{align}
P_{+}
={}&
\langle +|
\hat\rho_{\mathrm{S}}
|+\rangle
\notag\\
={}&
Z^{\varnothing}\cos^2\theta
+
P_{\mathrm B}\sin^2\theta
\notag\\
&+
\sqrt N\sin\theta\cos\theta
\left(
X_0^{\varnothing}
+
Y_0^{\varnothing}
\right),
\label{eq:Pplus-uvar}
\\
P_{-}
={}&
\langle -|
\hat\rho_{\mathrm{S}}
|-\rangle
\notag\\
={}&
Z^{\varnothing}\sin^2\theta
+
P_{\mathrm B}\cos^2\theta
\notag\\
&-
\sqrt N\sin\theta\cos\theta
\left(
X_0^{\varnothing}
+
Y_0^{\varnothing}
\right).
\label{eq:Pminus-uvar}
\end{align}
\end{subequations}
All of these observables are evaluated directly from the five unique variables of the physical representative. The dense $(N+1)\times(N+1)$ reduced density matrix is never reconstructed.

When the finite bath correlation decomposition contains $m>1$ exponential terms, each molecule carries a local hierarchy vector $\boldsymbol\alpha_i=(n_{i0},\ldots,n_{i,m-1})$ rather than a scalar occupation. A molecular relabeling moves this complete vector without changing the order of its exponential channels, and canonical representatives are formed from repeated local hierarchy vectors. The same unique-variable construction, grouped Hamiltonian sums, aligned target accessors, and matrix-free propagation therefore apply. At fixed $m$ and $L$, the canonical-pattern count still saturates for $N\geq L$, and the unique-variable dimension saturates for $N\geq L+2$. The number of retained patterns can nevertheless grow rapidly with both $m$ and $L$. The complete derivation, worked example, and numerical illustration for a multi-exponential decomposition are provided in the Supporting Information.

\section{Initial conditions and distinguished molecules}
\label{sec:initial}

\noindent The symmetry reduction used during propagation must also be consistent with the initial hierarchy. We first consider factorized system-bath initial conditions, then discuss localized initial states that distinguish particular molecules, and finally show how arbitrary initial density operators can be treated using linearity.

\subsection{Permutation-symmetric initial states}

\noindent For a factorized initial system-bath state, the physical zeroth-tier ADO contains the chosen system density operator, while all higher-tier ADOs initially vanish:
\begin{subequations}
\begin{align}
\hat R_{\varnothing}(0)
&=
\hat\rho_{\mathrm{S}}(0),
\label{eq:factorized-initial-zero}
\\
\hat R_{\Lambda}(0)
&=
0,
~~
|\Lambda|>0.
\label{eq:factorized-initial-higher}
\end{align}
\end{subequations}
The vanishing higher-tier ADOs automatically satisfy every molecular relabeling symmetry. The only remaining question is whether the initial physical density operator is unchanged by molecular relabeling:
\begin{equation}
\hat U_{\pi}
\hat\rho_{\mathrm{S}}(0)
\hat U_{\pi}^{\dagger}
=
\hat\rho_{\mathrm{S}}(0).
\label{eq:symmetric-initial-density}
\end{equation}
When this condition holds, the initial state can be represented directly using the fully compressed variables of the empty pattern.

The cavity population, bright state population, and polariton populations all satisfy Eq.~\ref{eq:symmetric-initial-density}. For example, the initial cavity population
\begin{equation}
\hat\rho_{\mathrm{S}}(0)
=
|c\rangle\langle c|
\end{equation}
is represented by
\begin{equation}
Z^{\varnothing}(0)=1,
~~
X_0^{\varnothing}(0)
=
Y_0^{\varnothing}(0)
=
P_0^{\varnothing}(0)
=
T_0^{\varnothing}(0)
=
0.
\label{eq:cavity-initial-uvar}
\end{equation}
For an initial bright state population,
\begin{equation}
\hat\rho_{\mathrm{S}}(0)
=
|\mathrm B\rangle\langle\mathrm B|,
\end{equation}
every molecular matrix element is equal. The corresponding compressed variables are
\begin{equation}
P_0^{\varnothing}(0)
=
T_0^{\varnothing}(0)
=
\frac{1}{N},
~~
Z^{\varnothing}(0)
=
X_0^{\varnothing}(0)
=
Y_0^{\varnothing}(0)
=
0.
\label{eq:bright-initial-uvar}
\end{equation}
Using the polariton convention in Eq.~\ref{eq:polariton-states}, an initial upper-polariton population is represented by
\begin{subequations}
\begin{align}
Z^{\varnothing}(0)
&=
\cos^2\theta,
\\
X_0^{\varnothing}(0)
=
Y_0^{\varnothing}(0)
&=
\frac{\sin\theta\cos\theta}{\sqrt N},
\\
P_0^{\varnothing}(0)
=
T_0^{\varnothing}(0)
&=
\frac{\sin^2\theta}{N}.
\end{align}
\label{eq:upper-polariton-initial}
\end{subequations}
\noindent Similarly, an initial lower-polariton population is represented by
\begin{subequations}
\begin{align}
Z^{\varnothing}(0)
&=
\sin^2\theta,
\\
X_0^{\varnothing}(0)
=
Y_0^{\varnothing}(0)
&=
-\frac{\sin\theta\cos\theta}{\sqrt N},
\\
P_0^{\varnothing}(0)
=
T_0^{\varnothing}(0)
&=
\frac{\cos^2\theta}{N}.
\end{align}
\label{eq:lower-polariton-initial}
\end{subequations}
\noindent Thus, cavity, bright, and polariton initial populations require only the five unique variables of the empty pattern.

\subsection{Localized initial states and distinguished molecules}

\noindent A localized molecular population,
\begin{equation}
\hat\rho_{\mathrm{S}}(0)
=
|e_1\rangle\langle e_1|,
\label{eq:localized-initial-state}
\end{equation}
is not unchanged by a permutation that exchanges molecule $1$ with another molecule. Molecule $1$ must therefore remain explicitly identifiable during the propagation.

The state is nevertheless unchanged by any relabeling that leaves molecule $1$ fixed and permutes only molecules $2,\ldots,N$:
\begin{equation}
\hat U_{\pi}
|e_1\rangle\langle e_1|
\hat U_{\pi}^{\dagger}
=
|e_1\rangle\langle e_1|,
~~
\pi(1)=1.
\label{eq:localized-residual-symmetry}
\end{equation}
Consequently, only molecule $1$ must be treated as distinguished. Its system row, system column, and local hierarchy label remain explicit. The hierarchy labels of molecules $2,\ldots,N$ can still be sorted and compressed exactly as before.

More generally, suppose an initial operator distinguishes $d$ molecular labels. These $d$ labels are kept explicit, while the remaining $N-d$ molecules are treated as equivalent and their local hierarchy labels are sorted canonically. The effective size of the compressible molecular ensemble is therefore
\begin{equation}
N_{\mathrm{sym}}=N-d.
\label{eq:N-symmetric-pool}
\end{equation}

The canonical-pattern count saturates when the compressible part contains at least $L$ molecules:
\begin{equation}
N-d\geq L.
\end{equation}
The unique variable dimension saturates when it contains at least $L+2$ molecules:
\begin{equation}
N-d\geq L+2.
\end{equation}
Equivalently, the saturation conditions are
\begin{subequations}
\begin{align}
N &\geq L+d ~~~~~~~ \rightarrow
~~~~~
\text{canonical pattern},
\label{eq:distinguished-pattern-threshold}
\\
N &\geq L+d+2 ~ ~\rightarrow
~~~~~
\text{unique variables}.\label{eq:distinguished-thresholds}
\end{align}
\end{subequations}
For a single localized excitation, $d=1$, and these conditions become
\begin{subequations}
\begin{align}
N &\geq L+1 ~~ \rightarrow
~~~~~
\text{canonical pattern},
\\
N &\geq L+3 ~~\rightarrow
~~~~~
\text{unique variables}.
\end{align}
\end{subequations}
A localized initial state therefore enlarges the compressed representation relative to a fully permutation-symmetric initial state. However, it does not restore the combinatorial growth of the molecule-resolved hierarchy because the remaining $N-1$ molecules are still treated collectively.

\subsection{General initial density operators}

\noindent The HEOM evolution is linear in the initial density operator. We denote the reduced dynamical map by $\hat{\mathcal G}_t$, such that
\begin{equation}
\hat\rho_{\mathrm{S}}(t)
=
\hat{\mathcal G}_t
\Big[
\hat\rho_{\mathrm{S}}(0)
\Big].
\label{eq:Gt}
\end{equation}
For two initial operators $\hat A$ and $\hat B$ and complex coefficients $a$ and $b$, linearity gives
\begin{equation}
\hat{\mathcal G}_t
\left[
a\hat A+b\hat B
\right]
=
a\hat{\mathcal G}_t[\hat A]
+
b\hat{\mathcal G}_t[\hat B].
\label{eq:Gt-linearity}
\end{equation}

\noindent An arbitrary initial operator in the single excitation manifold can be expanded as
\begin{align}
\hat\rho_{\mathrm{S}}(0)
={}&
\rho_{cc}(0)|c\rangle\langle c|
\notag\\
&+
\sum_{i=1}^{N}
\rho_{ic}(0)|e_i\rangle\langle c|
+
\sum_{i=1}^{N}
\rho_{ci}(0)|c\rangle\langle e_i|
\notag\\
&+
\sum_{i=1}^{N}
\rho_{ii}(0)|e_i\rangle\langle e_i|
\notag\\
&+
\sum_{\substack{i,j=1\\i\neq j}}^{N}
\rho_{ij}(0)|e_i\rangle\langle e_j|.
\label{eq:arbitrary-initial-expansion}
\end{align}
The five terms of Eq.~\ref{eq:arbitrary-initial-expansion} contain five distinct ordered operator types: a cavity population, a molecule-cavity coherence, a cavity-molecule coherence, a molecular population, and an ordered intermolecular coherence.

Molecular covariance of the dynamical map gives
\begin{equation}
\hat{\mathcal G}_t \left[ \hat U_{\pi}\hat{A}\hat U_{\pi}^{\dagger} \right] = \hat{U}_{\pi} \hat{\mathcal{G}}_t \big[\hat A\big] \hat{U}_{\pi}^{\dagger}.
\label{eq:Gt-covariance}
\end{equation}
Therefore, the response to any molecularly relabeled operator can be obtained from the response to one representative operator of the same type. For example, if $\pi(1)=i$, then
\begin{equation}
\hat{\mathcal G}_t
\Big[
|e_i\rangle\langle c|
\Big]
=
\hat U_{\pi}
\hat{\mathcal G}_t
\Big[
|e_1\rangle\langle c|
\Big]
\hat U_{\pi}^{\dagger}.
\label{eq:reconstruct-ec}
\end{equation}
Similarly, if $\pi(1)=i$ and $\pi(2)=j$, then
\begin{equation}
\hat{\mathcal G}_t
\Big[
|e_i\rangle\langle e_j|
\Big]
=
\hat U_{\pi}
\hat{\mathcal G}_t
\Big[
|e_1\rangle\langle e_2|
\Big]
\hat U_{\pi}^{\dagger},
~~ i\neq j.
\label{eq:reconstruct-ee}
\end{equation}
It follows that the evolution of an arbitrary initial operator can be reconstructed from the representative propagations summarized in Table~\ref{tab:representative-initial-operators}.
\begin{table*}
\centering
\setlength{\tabcolsep}{10pt}
\renewcommand{\arraystretch}{1.0}
\begin{tabular}{|c|c|c|c|}
\hline
Representative & Interpretation & Distinguished labels & Possible adjoint reduction\\
\hline
$|c\rangle\langle c|$ & Cavity population & $0$ & Not applicable\\
$|e_1\rangle\langle c|$ & Molecule-cavity coherence & $1$ & Paired with $|c\rangle\langle e_1|$\\
$|c\rangle\langle e_1|$ & Cavity-molecule coherence & $1$ & Paired with $|e_1\rangle\langle c|$\\
$|e_1\rangle\langle e_1|$ & Localized molecular population & $1$ & Not applicable\\
$|e_1\rangle\langle e_2|$ & Ordered intermolecular coherence & $2$ & Its adjoint follows same relabeling
\\
\hline
\end{tabular}
\caption{Representative initial system operators required for linear reconstruction. The number of distinguished labels specifies how many molecular identities remain explicit during the corresponding propagation. The two ordered cavity-molecule coherences can be related by an adjoint operation only when the bath decomposition, hierarchy truncation, and closure preserve the required conjugation relation.}
\label{tab:representative-initial-operators}
\end{table*}

If the reduced map preserves the adjoint operation,
\begin{equation}\label{eq:adjoint-preservation}
\hat{\mathcal G}_t
\big[
\hat A^{\dagger}
\big]
=
\hat{\mathcal G}_t \big[\hat A \big]^{\dagger},
\end{equation}
the response to $|c\rangle\langle e_1|$ can be obtained by taking the adjoint of the response to $|e_1\rangle\langle c|$. In that case, only four independent representative propagations are required. For bath decompositions containing complex decay rates $\nu_k$, Eq.~\ref{eq:adjoint-preservation} must not be assumed automatically. It holds only when the complex exponential channels occur in the required conjugate pairs and the hierarchy truncation and closure preserve the corresponding conjugation relation. Without these conditions, the two ordered cavity-molecule coherences should be propagated separately.

The number of representative propagations needed to describe an arbitrary initial operator is therefore independent of $N$. However, if one explicitly requests every entry of the final dense density matrix, the reconstruction must produce $(N+1)^2$ matrix elements and consequently requires $O(N^2)$ output storage. This quadratic cost belongs to the requested dense output itself, not to the compressed HEOM propagation.

\section{Results and discussion}
\label{sec:results}
\noindent We use the numerical examples below to verify the symmetry-adapted propagation and to illustrate how the population dynamics change with molecular ensemble size. For all the simulations presented below, the molecule and cavity excitation energies are taken to be, $\omega_{\mathrm{cav}} = \epsilon_{\mathrm{x}} = 10000$ cm$^{-1}$ and the the Rabi splitting is taken to be $\Omega_{\mathrm{R}} = 500$ cm$^{-1}$. The single molecule light-matter coupling changes with $N$ according to Eq.~\ref{eq:Rabi-splitting},
\begin{equation}
g(N)=\frac{\Omega_{\mathrm R}}{2\sqrt{N}}.
\label{eq:g-fixed-Rabi}
\end{equation}
Consequently, the resonant UP and LP energies remain unchanged as $N$ varies, while the coupling of any individual molecule to the cavity
decreases as $N^{-1/2}$. For the one-exponential decomposition per local bath, the retained Drude-Lorentz contribution is generated from
\begin{equation}\label{eq:Drude_Lorentz}
J_{\mathrm{DL}}(\omega)
=
\frac{2\lambda_{\mathrm{DL}}\gamma_{\mathrm{DL}}\omega}
{\omega^2+\gamma_{\mathrm{DL}}^2},
\end{equation}
where $\lambda_{\mathrm{DL}}$ is the reorganization energy and $\gamma_{\mathrm{DL}}$ is the bath decay rate. We use $\lambda_{\mathrm{DL}}=50\,\mathrm{cm}^{-1}$, $\gamma_{\mathrm{DL}}=18\,\mathrm{cm}^{-1}$, and $T=300~\mathrm{K}$, for which $k_{\mathrm B}T\approx208~\mathrm{cm}^{-1}$. The residual contribution from the omitted Matsubara terms is treated with the Ishizaki-Tanimura terminator.~\cite{Ishizaki2005JPSJ} Unless stated otherwise, the symmetry adapted calculations use hierarchy depth $L=25$ and time step $\Delta t=0.5\,\mathrm{fs}$. The $N=2$ comparisons with the QuTiP~\cite{Lambert2023PRR,Lambert2026PR} implementation of conventional HEOM use $L=15$.

\subsection{Single exponential bath}

\noindent We now demonstrate the ability of the unique variable reduced HEOM method to simulate large polaritonic systems. We choose the upper polariton (UP) state, defined in Eq.~\ref{eq:polariton-states}a, as the initial state of the system. On resonance, the initially populated upper polariton contains equal cavity and bright molecular contributions,
\begin{equation}
|+\rangle
=
\frac{1}{\sqrt{2}}
\Big(
|c\rangle+|\mathrm B\rangle
\Big).
\label{eq:resonant-UP-composition}
\end{equation}

\begin{figure}
    \centering
    \includegraphics[width=\linewidth]{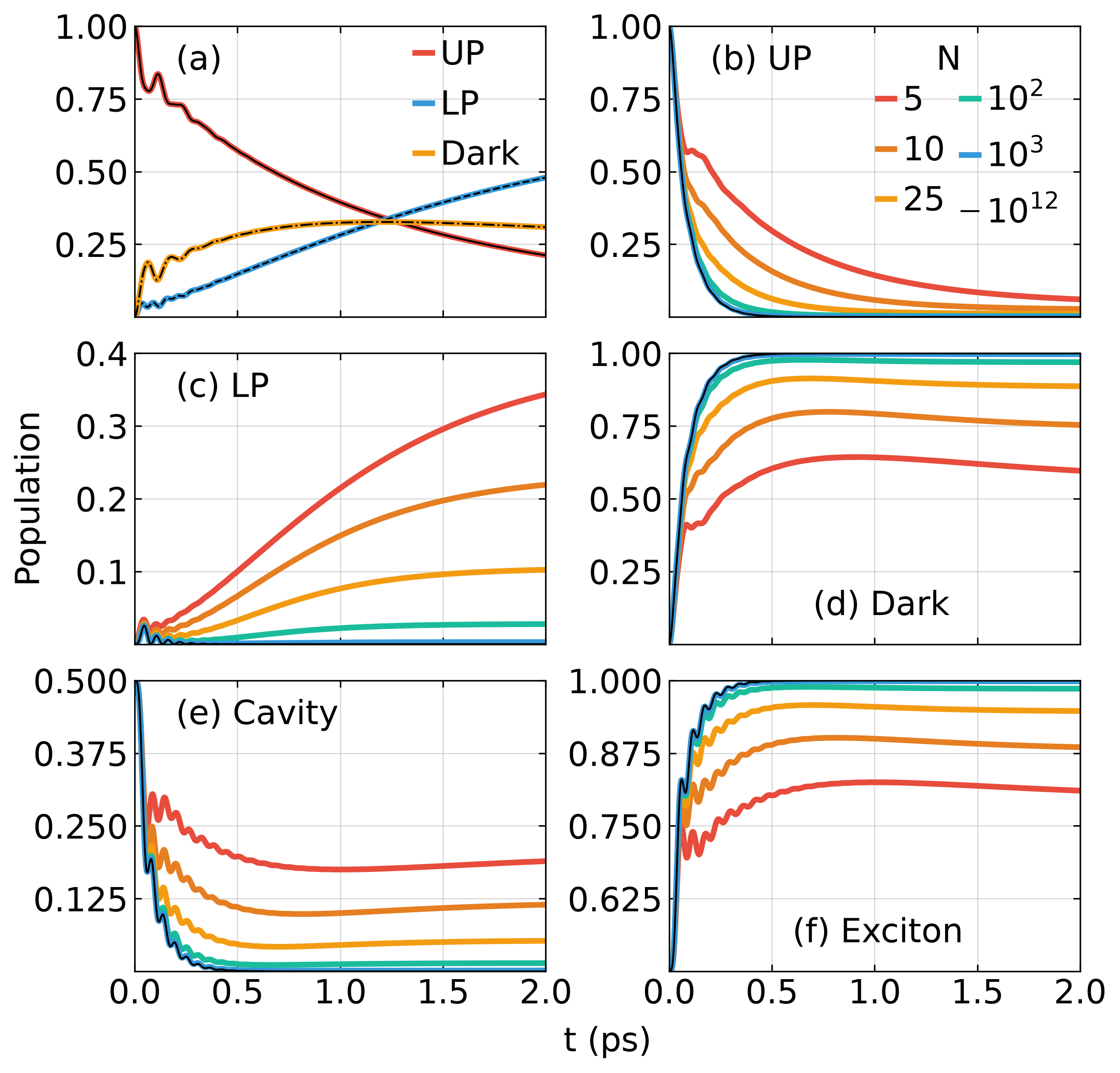}
    \caption{Population dynamics for an initial upper-polariton population with a one-exponential decomposition per local bath. (a) Comparison between symmetry-adapted HEOM and the QuTiP implementation of conventional HEOM for $N=2$. Colored curves show the symmetry-adapted calculation, and the overlaid black traces show the corresponding conventional HEOM results for the UP, LP, and total dark populations. (b) UP, (c) LP, (d) total dark, (e) cavity, and (f) total exciton populations for $N=5$, $10$, $25$, $10^2$, $10^3$, and $10^{12}$, as identified by the legend.} \label{fig:varyN_oneExp_population}
\end{figure}

In Fig.~\ref{fig:varyN_oneExp_population}(a), we compare the symmetry-adapted unique variable calculation with conventional HEOM for $N=2$ system. The UP, LP, and total dark state populations, which are represented by the red, blue, and yellow curves respectively, fully agree with the conventional HEOM simulations of QuTiP, which are represented by the corresponding black curves in the figure. This agreement confirms that the canonical-ADO and unique variable reductions preserve the dynamics of the corresponding finite HEOM calculation.

In Fig.~\ref{fig:varyN_oneExp_population}(b)-(d) we present the UP, LP and the total dark-state population dynamics by varying the $N$ at the same Rabi splitting of $\Omega_{\mathrm{R}} = 500$ cm$^{-1}$. For $N=5$, represented by the red curves, there is an initial fast decay of the UP state upto $\approx 100$~fs, which is accompanied by the fast increase of the Dark state population. The LP populations does not significantly increase at this stage. The UP population then follows a slower decay, and we see a slow increment in the dark state as well as the LP populations. Around 700 fs, the Dark state starts decaying slowly and transferring more population to the LP. For $N=10$, represented by the orange curves in the figure, we see a similar behaviour of the UP, LP, and dark state populations. The second decay of the UP population is faster than in the $N=5$ case and there is also an overall decrease in the amount of population transfer from the dark to LP state. With increasing $N$ value, the second decay behaviour of the UP population slowly disappears and for $N=100$ (represented by the green curve), there is no longer any visible second decay timescale. At $N=100$, there is still a small finite population transfer to the LP state from the dark state. The dynamics for $N=10^3$ and $N=10^{12}$, represented by light and dark blue curves respective, have no visible difference in the timescale considered and we can safely assume that for these parameters, the system has reached the thermodynamic limit, with the LP state only acquiring populations transiently and coherently during the very early stages of the dynamics (upto roughly 250~fs) and all the population transfers to dark states. At this limit, the UP and dark state populations represent a simple kinetic rate model kind of population transfer.~\cite{Lai2024JCP, Lai2026JCP}.

In Fig.~\ref{fig:varyN_oneExp_population}(e)-(f), we present the population dynamics for the cavity and total exciton states, respectively, when the initial condition is the UP state and thus both the cavity and the exciton have 0.5 population at $t=0~$fs. For $N=5$ case, we see an initial fast decay of the cavity populations to the excitons up to 100~fs. After 100~fs, the rate of transfer slows down, and around 700~fs, there is a reversal of net population flow, causing the cavity population to increase. For $N=10$ and $N=25$, we see a similar behavior, but the overall magnitude of the cavity population is reduced. At $N=100$, we no longer observe the revival behavior of the cavity population. At $N=10^3$, the system reached the thermodynamic limit, and the behavior converges, with all population getting transferred to the excitons. It is to be noted that even at $N=10^{12}$, the initial transient population dynamics still shows coherent oscillations.

\subsection{Dynamics from a localized molecular excitation}
\begin{figure}[H]
    \centering
    \includegraphics[width=\linewidth]{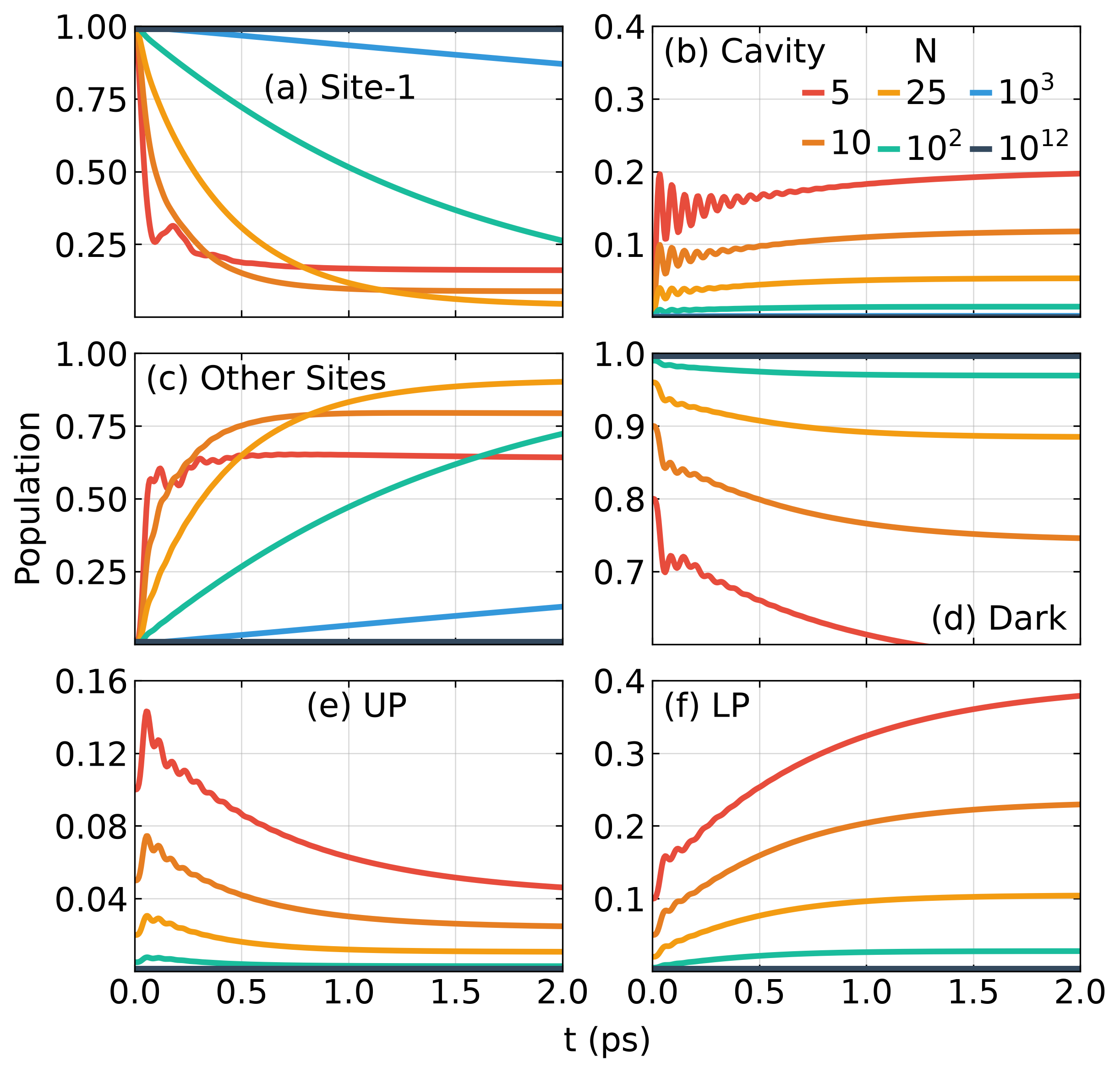}
    \caption{Population dynamics with unique variable HEOM for the localized initial condition of Eq.~\ref{eq:localized_initial_condition}. Panels (a)-(c) present the populations of the first site, cavity and the total population of other sites, respectively, for different number of molecules ($N$) coupled to cavity. Panels (d)-(f) present the total dark populations, and the populations of the upper and lower polariton states respectively, for different number of molecules ($N$) coupled to cavity. In all the panels, the different $N = 5, 10, 25, 10^2, 10^3, 10^{12}$ cases are represented by solid line curves with red, orange, yellow, green, blue and dark blue respectively.} \label{fig:varyN_oneExp_localized_population}
\end{figure}

\noindent We next consider the nonsymmetric initial condition
\begin{equation}
\hat\rho_{\mathrm{S}}(0)
=
|e_1\rangle\langle e_1|.
\label{eq:localized_initial_condition}
\end{equation}
This propagation retains molecule $1$ as a distinguished label while compressing the remaining $N-1$ molecules. 

In Fig.~\ref{fig:varyN_oneExp_localized_population}(a)-(c), we present the site-1 (initially localized excitation), cavity and total site (excluding site-1) population dynamics for varying $N$. For $N=5$ system (the red curves in all panels), there is an initial fast decay of the localized site-1 population which is accompanied by initial coherent increase of populations in cavity and other sites as can be seen in panels (b) and (c). After the initial transient period, the populations reach a steady state. The steady cavity population is approximately 0.2 $\simeq 1/5$. The $N=10$ and $N=25$ systems (represented by the orange and yellow curves respectively in all panels) show a similar saturation behaviour where the steady state cavity populations show a $1/N$ behaviour. Unlike $N=5$ system, the $N=10$ and $N=25$ systems do not show two seperate timescale behaviour in the decay and with increase in $N$, the rate of decay of the site-1 populations decrease. For $N=100$, there is almost no population in the cavity mode and almost all of the population is redistributed to the other sites. For $10^3$ molecules the rate of redistribution is much lower and for $N=10^{12}$, there is no visible transfer to other states within the 2~ps time window and a localized excitations stays localized without any redistribution.

On resonance, the localized state contains bright state weight $1/N$ and dark state weight $1-1/N$. The initial TC-basis (Eq.~\ref{eq:polariton-states}) populations are therefore
\begin{subequations}
\begin{gather}\label{eq:localized-initial-TC-populations}
P_{+}(0)
=
P_{-}(0)
=
\frac{1}{2N},
\\
P_{\mathrm D}(0)
=
1-\frac{1}{N}.
\end{gather}
\end{subequations}
The nonzero initial polariton populations arise from the projection of $|e_1\rangle$ onto the collective bright state. In Fig.~\ref{fig:varyN_oneExp_localized_population}(d)-(f) we present the TC-eigenbasis populations which provide a complementary interpretation. For $N=5$, Eq.~\ref{eq:localized-initial-TC-populations} gives an initial dark population of $0.8$ and initial UP and LP populations of $0.1$ each. There is a fast decay of the dark state population in the first 100~fs to the LP and UP states which is followed by a slower decay. The inital spike in the UP population is followed by a decay back to the dark states. Most of the dark state decay is to the LP state at later stage of dynamics. With increaing $N$, the initial polariton participation decreases as $1/N$, while the dark contribution approaches unity. The subsequent changes in the UP and LP populations are therefore progressively reduced with increasing $N$. At higher $N$ values, the population gets trapped in the dark subspace with no tranfers to the brigh polariton manifold.  Together, the site and TC-eigenbasis representations demonstrate that distinguished molecule propagation retains both the localized component and its collective bright state contribution without restoring the full molecule-resolved hierarchy.

\subsection{Dynamics from a coherent cavity-site superposition}

\noindent The final single exponential example uses the coherent nonsymmetric initial state
\begin{subequations}
\begin{gather}
|\Psi_{\mathrm{S}}(0)\rangle
=
\sqrt{0.3}\,|e_1\rangle
+
\sqrt{0.7}\,|c\rangle, \label{eq:site1_cavity_superposition}
\\
\hat\rho_{\mathrm{S}}(0)
=
|\Psi_{\mathrm{S}}(0)\rangle
\langle\Psi_{\mathrm{S}}(0)|.
\label{eq:site1_cavity_superposition_density}
\end{gather}    
\end{subequations}
Its nonzero populations and coherences are
\begin{subequations}\label{eq:site1-cavity-initial-elements}
\begin{align}
\langle e_1|\hat\rho_{\mathrm{S}}(0)|e_1\rangle
&=
0.3,
\\
\langle c|\hat\rho_{\mathrm{S}}(0)|c\rangle
&=
0.7,
\\
\langle e_1|\hat\rho_{\mathrm{S}}(0)|c\rangle
=
\langle c|\hat\rho_{\mathrm{S}}(0)|e_1\rangle
&=
\sqrt{0.21}.
\end{align}
\end{subequations}
\begin{figure}
    \centering
    \includegraphics[width=\linewidth]{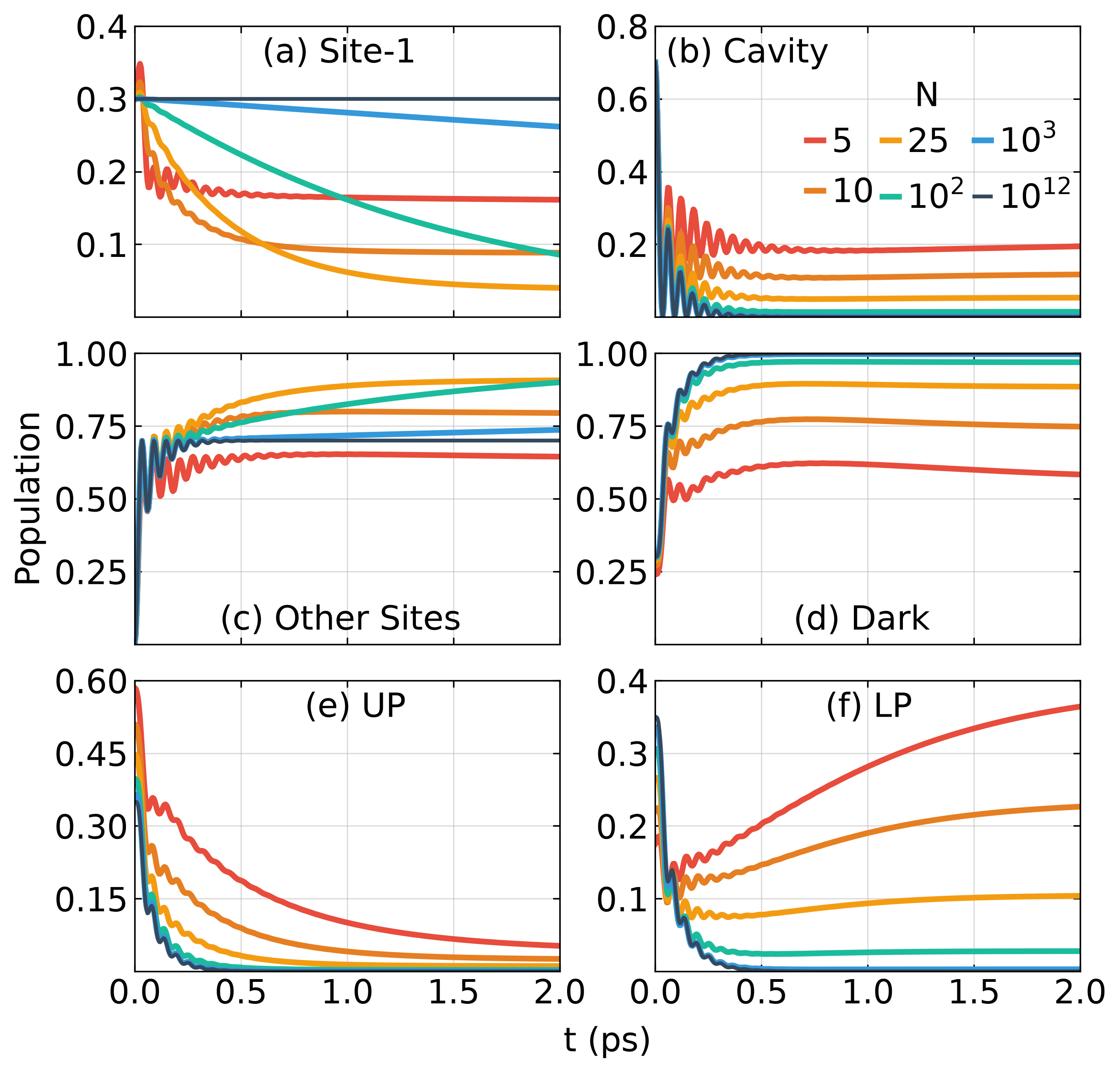}
    \caption{Population dynamics for the cavity-site superposition initial condition in Eq.~\ref{eq:site1_cavity_superposition}. Panels (a), (b), and (c) show the population of molecule $1$, the cavity population, and the total population of the remaining molecular sites, respectively. Panels (d), (e), and (f) show the total dark state manifold, UP, and LP populations. Curves correspond to $N=5$, $10$, $25$, $10^2$, $10^3$, and $10^{12}$, as identified by the legend.}
    \label{fig:varyN_oneExp_cavity_site_superposition}
\end{figure}
In Fig.~\ref{fig:varyN_oneExp_cavity_site_superposition}(a)-(c) we present the site-1, cavity and total site (excluding site-1) population dynamics for different $N$ values for the initial condition of Eq.~\ref{eq:site1_cavity_superposition}. For $N = 5$, represented by the red curve in all panels, there is an initial fast decay of the site-1 and cavity populations to the other sites. After this intial transfer to the sites, there is coherent exchange of populations between the cavity and the sites, following which there is no further transfer and the cavity population reaches a steady state of approximately 0.2 ($\simeq 1/5$). For increasing $N$, we see a similar initially coherent transfer from the cavity to the other sites with the cavity steady state population showing a $1/N$ behaviour. For the site-1, with increasing $N$, the rate of decay of population slows down and at $N=10^{12}$ we see that there is no transfer of population from site-1 and that it reamins at its initial population of 0.3. The increase of the total population of the other sites shows a similar behaviour as in the $N=5$ case and with increase in $N$, the increase of the other site populations comes mostly from the cavity decay. At $N=10^{12}$, all of the cavity population is transferred to the other sites and there is no exchange with the site-1 population.

Using the UP and LP convention in Eq.~\ref{eq:polariton-states}, the resonant initial TC-eigenbasis populations are
\begin{subequations}
\begin{align}
P_{\mathrm D}(0)
&=
0.3\left(1-\frac{1}{N}\right),
\\
P_{\pm}(0)
&=
\frac{1}{2}
\left(
\sqrt{0.7}
\pm
\sqrt{\frac{0.3}{N}}
\right)^2,
\end{align}
\label{eq:superposition-initial-TC-populations}
\end{subequations}

The different initial UP, LP, and dark populations are therefore fixed by the coherent cavity-bright projection of the prepared state and vary explicitly with $N$. 

\begin{figure*}[htbp]
    \centering \includegraphics[width=0.85\linewidth]{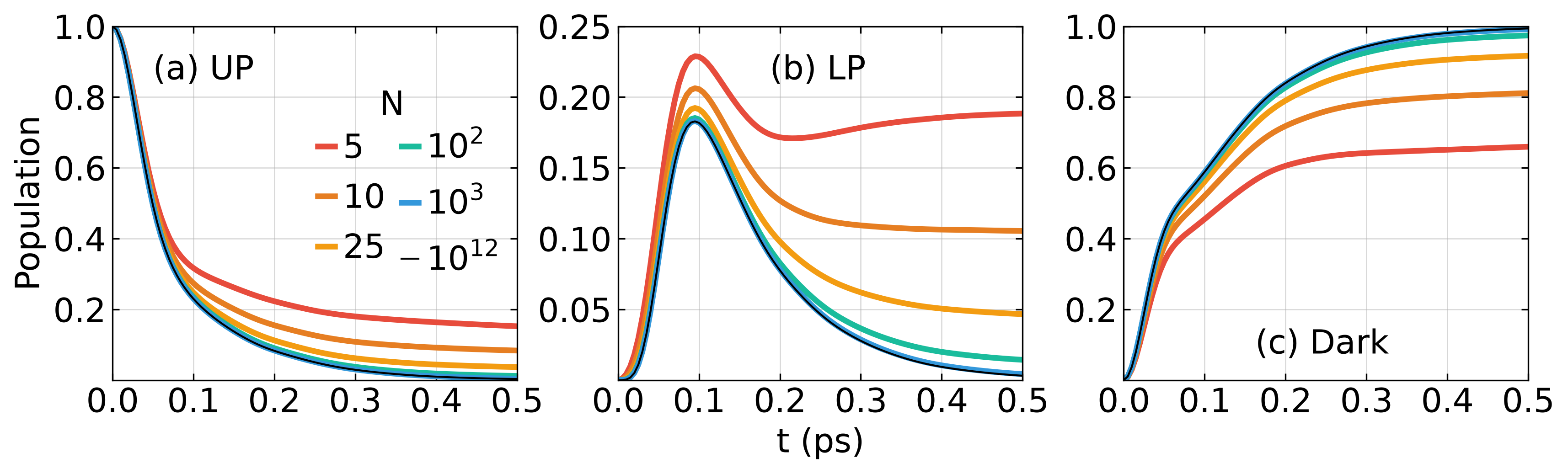} \caption{Disorder-averaged population dynamics following initialization in the upper polariton of the mean resonant HTC Hamiltonian in the presence of independent Gaussian static energetic disorder and identical local Drude-Lorentz environments. Panels show the populations of the reference (a) upper polariton, (b) lower polariton, and (c) total dark manifold for $N=5$, $10$, $25$, $10^2$, $10^3$, and $10^{12}$. The reference states are defined by the disorder-free TC Hamiltonian.} \label{fig:varyN_static}
\end{figure*}

In Fig.~\ref{fig:varyN_oneExp_cavity_site_superposition}(d)-(f) we present the total dark state, UP and LP population dynamics for different $N$ values for the initial condition of Eq.~\ref{eq:site1_cavity_superposition}. The initial values in Eq.~\ref{eq:superposition-initial-TC-populations} account for the ensemble-size dependent intercepts in Fig.~\ref{fig:varyN_oneExp_cavity_site_superposition}(d)-(f). For smaller ensembles, the localized molecular component has a larger bright state projection, producing a stronger imbalance between the initial UP and LP populations. As $N$ increases, this molecular bright component decreases, $P_{\mathrm D}(0)$ approaches $0.3$, and both polariton populations approach $0.35$.

The polariton states in Fig.~\ref{fig:varyN_oneExp_cavity_site_superposition}(d)-(f) show very similar behavious as observed in Fig.~\ref{fig:varyN_oneExp_population}. For small $N$ values there are two timescales of decay of the UP population: an initial fast decay followed by a slower decay and with increase in $N$, the second decay mechanism disappears and we have an overall fast decay of UP state. The LP state on the other hand, at small $N$ values, shows an increase of population after its initial decrease. With increase of $N$ value, this behaviour disappears and at $N=10^3$ the population of LP only decays. The decay of UP and LP states is accompanied by the increase of dark state populations and the amount of population transferred to dark states increases with $N$ with the total dark state population becoming 1 at long times for $N=10^3$. It is to be noted that even at $N=10^{12}$, unlike Fig.~\ref{fig:varyN_oneExp_population}, the initial population dynamics of polariton states show coherent oscillations.

\subsection{Static Energetic Disorder}
\noindent We now present the polariton dynamics in presence of diagonal inhomogeneous static disorder~\cite{Mondal2025JCP1, li2025JCP, li2026arXiv}. The TC-system Hamiltonian in Eq.~\ref{eq:HS} is modified as,
\begin{equation}\label{eq:HS_disorder}
\hat H_{\mathrm S}(\boldsymbol{\delta})
=
\hat H_{\mathrm S}
+
\sum_{i=1}^{N}\delta\epsilon_i\hat{Q}_i,
~~~~
\hat{Q}_i=|e_i\rangle\langle e_i|.
\end{equation}
The exciton diagonal energies is sampled from an inhomogeneous gaussian distribution,
\begin{equation}\label{eq:static-distribution}
P(\boldsymbol{\delta})
=
\prod_{i=1}^{N}
\frac{1}{\sqrt{2\pi\sigma^2}}
\exp\left(-\frac{\delta\epsilon_i^2}{2\sigma^2}\right).
\end{equation}
Since the distribution is static and follows a gaussian statistics, we can represent the diagonal disorder as a diagonal bath with a non-decaying correlation function~\ref{eq:BCF-spectral},
\begin{equation}
    C_{\mathrm{static}}(t) = \sigma^2 \cdot e^{-0\cdot t} = \sigma^2
\end{equation}
This adds an HEOM bath channel to every molecule with the expansion coefficient $\sigma^2$ and decay rate of 0.

Fig.~\ref{fig:varyN_static} shows the disorder-averaged dynamics following initialization in the upper polariton (Eq.~\ref{eq:UP}) of the mean resonant TC Hamiltonian (Eq.~\ref{eq:HS}). We use $\omega_{\mathrm{cav}} = \epsilon_{\mathrm x} = 10000~\mathrm{cm}^{-1}$, $\Omega_{\mathrm R} = 100~\mathrm{cm}^{-1}$, $\sigma = 25~\mathrm{cm}^{-1}$. The collective Rabi splitting is held fixed as $N$ is varied according to Eq.~\ref{eq:Rabi-splitting}. In addition to the static channel, each molecule is coupled to an identical Drude-Lorentz environment with $\lambda_{\mathrm{DL}} = 50~\mathrm{cm}^{-1}$, $\gamma_{\mathrm{DL}} = 18~\mathrm{cm}^{-1}$, $T=300~\mathrm K$ for which $k_{\mathrm B}T\simeq208~\mathrm{cm}^{-1}$. We used the multi-exponential formalism to represent the two channels (Drude-Lorentz and static disorder) where the total hierarchy depth converged at $L=14$ for the above parameters. The details of multi-exponential formalism of symmetry-adapted HEOM and the static disorder is discussed in the Supplementary Information.

The static energy fluctuations mix the permutation-symmetric bright state with the molecular dark manifold in each realization. After disorder averaging, this appears as rapid loss of bright-state coherence and growth of the total dark population. The dynamical Drude-Lorentz environments act simultaneously and allow energy relaxation in addition to the inhomogeneous dephasing generated by the static channel.

\begin{table*}[t]
\centering
\setlength{\tabcolsep}{10pt}
\renewcommand{\arraystretch}{1.0}
\begin{tabular}{|c|c|c|c|c|c|c|}
\hline
$N$
&
\shortstack{$\mathcal{N}_{\mathrm{ADO}}^{\mathrm{naive}}$}
&
\shortstack{$\mathcal{M}_{\mathrm{ADO}}^{\mathrm{naive}}$ (GB)}
&
\shortstack{$\mathcal{N}_{\mathrm{ADO}}^{\mathrm{can}}$}
&
\shortstack{$\mathcal{N}_{\mathrm{var}}^{\mathrm{unq}}$}
&
\shortstack{$\mathcal{M}_{\mathrm{var}}^{\mathrm{unq}}$(GB)}
&
\shortstack{ $\mathcal{R}$-factor}
\\
\hline
$5$
& $1.43\times10^{5}$
& $8.21\times10^{-2}$
& $2.60\times10^{3}$
& $6.21\times10^{4}$
& $9.94\times10^{-4}$
& $8.26\times10^{1}$
\\
$10$
& $1.84\times10^{8}$
& $3.55\times10^{2}$
& $7.53\times10^{3}$
& $2.44\times10^{5}$
& $3.91\times10^{-3}$
& $9.10\times10^{4}$
\\
$25$
& $1.26\times10^{14}$
& $1.37\times10^{9}$
& $9.30\times10^{3}$
& $3.07\times10^{5}$
& $4.91\times10^{-3}$
& $2.79\times10^{11}$
\\
$10^{2}$
& $1.30\times10^{26}$
& $2.12\times10^{22}$
& $9.30\times10^{3}$
& $3.07\times10^{5}$
& $4.91\times10^{-3}$
& $4.32\times10^{24}$
\\
$10^{3}$
& $8.90\times10^{49}$
& $1.43\times10^{48}$
& $9.30\times10^{3}$
& $3.07\times10^{5}$
& $4.91\times10^{-3}$
& $2.91\times10^{50}$
\\
$10^{12}$
& $6.45\times10^{274}$
& $1.03\times10^{291}$
& $9.30\times10^{3}$
& $3.07\times10^{5}$
& $4.91\times10^{-3}$
& $2.10\times10^{293}$
\\
\hline
\end{tabular}
\caption{
Analytical hierarchy sizes and raw state-vector storage for one exponential term per molecular bath and total hierarchy depth $L=25$. The two storage columns count one copy of the corresponding complex-double state and are not measurements of peak resident memory. The canonical-ADO and unique-variable columns refer to a fully permutation-symmetric propagation. The reduction factor $\mathcal{R}$ is the ratio of naive ADO storage to unique-variable
storage.
}
\label{tab:computational-resources-L25}
\end{table*}

For $N=5$, the UP population decreases rapidly, while the LP population develops a pronounced early-time maximum and remains appreciable over the displayed interval. The LP curve also exhibits a weak late-time upturn after its initial decrease. As $N$ increases, the remaining UP and LP populations are reduced and a progressively larger fraction of the excitation accumulates in the dark manifold. For the largest ensembles, the dark population approaches unity within the plotted time window, and the curves for $N=10^3$ and $N=10^{12}$ are nearly coincident. This agreement is a physical convergence with ensemble size for the chosen parameters and it is distinct from the earlier saturation of the computational representation with $N$.

\subsection{Comment on the computational resources utilized}
\label{sec:computational-resources}

\noindent Table~\ref{tab:computational-resources-L25} summarizes the analytical storage requirements for the one-exponential bath calculations at hierarchy depth $L=25$. The molecular ensemble sizes are the same as those used in the population dynamics figures. In the single excitation basis $\mathcal B_N=\{|c\rangle,|e_1\rangle,\ldots,|e_N\rangle\}$, every conventional ADO is a dense $(N+1)\times(N+1)$ complex matrix. For one exponential term per molecule, the conventional hierarchy and its raw storage are
\begin{subequations}
\begin{gather}
\mathcal N_{\mathrm{ADO}}^{\mathrm{naive}}(N,1,25)
=
\binom{N+25}{25},
\label{eq:resources-naive-count}
\\
\mathcal M_{\mathrm{ADO}}^{\mathrm{naive}}
=
16(N+1)^2~
\mathcal{N}_{\mathrm{ADO}}^{\mathrm{naive}}
~\text{bytes},
\label{eq:resources-naive-memory}
\end{gather}    
\end{subequations}
where each complex double-precision number occupies $16$ bytes. The unique-variable storage is
\begin{equation}
\mathcal M_{\mathrm{var}}^{\mathrm{unq}}
=
16~\mathcal{N}_{\mathrm{var}}^{\mathrm{unq}}
~\text{bytes}.
\label{eq:resources-uvar-memory}
\end{equation}
The memory values in Table~\ref{tab:computational-resources-L25} are reported in decimal gigabytes, with $1\,\mathrm{GB}=10^9$ bytes. The reduction factor reported below compares the complete number of complex
entries in the molecule-resolved hierarchy with the number of propagated
unique variables,
\begin{align}
\mathcal R(N)
=
\frac{
\mathcal N_{\mathrm{ADO}}^{\mathrm{naive}}(N,1,25)(N+1)^2
}{
\mathcal N_{\mathrm{var}}^{\mathrm{unq}}(N,1,25)
}
=
\frac{ \mathcal{M}_{\mathrm{ADO}}^{\mathrm{naive}}
}{
\mathcal{M}_{\mathrm{var}}^{\mathrm{unq}}
}.
\label{eq:resources-reduction-factor}
\end{align}

The conventional hierarchy becomes impractical at remarkably small ensemble sizes. At $N=10$, storing only one copy of all ADO matrices would already require $3.55\times10^{2}\,\mathrm{GB}$, whereas the corresponding unique-variable vector requires only $3.91\times10^{-3}\,\mathrm{GB}$, or $3.91\,\mathrm{MB}$. The storage reduction at this ensemble size is $9.10\times10^{4}$. At $N=25$, the conventional one-vector lower bound reaches $1.37\times10^{9}\,\mathrm{GB}$, while the unique-variable vector remains only $4.91\,\mathrm{MB}$, corresponding to a reduction by a factor of $2.79\times10^{11}$. The conventional estimates for the larger ensembles are astronomical and demonstrate that a direct molecule-resolved HEOM calculation is not merely expensive, but impossible on existing state-of-art computational hardware.

The two symmetry reductions remove different parts of this growth. At $L=25$, the canonical hierarchy contains $9296$ ADO representatives for every $N\geq25$, instead of the rapidly increasing number of molecule-labeled ADOs. If these representatives were retained as full matrices, however, their storage would still grow as $(N+1)^2$. The second reduction removes this remaining matrix redundancy. The fully symmetric hierarchy contains exactly $306743$ unique variables at $N=25$, $306750$ at $N=26$, and $306751$ for every $N\geq27$. Its raw state-vector storage therefore saturates at only $4.91\,\mathrm{MB}$, independent of any further increase in the number of molecules. The enormous calculation at $N=10^{12}$ consequently uses the same number of propagated variables as the calculation at $N=27$; only the multiplicity coefficients and the $N$-dependent system parameters change.

The memory values in Table~\ref{tab:computational-resources-L25} deliberately do not include the work arrays required by the time integrator. A conventional $4^{\mathrm{th}}$ Runge-Kutta implementation commonly retains the current state, four stage vectors, and one or more temporary or output vectors. Its state-sized working storage is therefore typically about six to seven times the one-vector values reported in the table, although the precise factor depends on the implementation and can be reduced with a low-storage integrator. 

Localized molecular excitation and cavity-site superposition dynamics each distinguish one molecule and therefore use the same partially symmetric hierarchy structure. At $L=25$, the corresponding canonical-pattern count saturates at $41391$ for $N\geq26$, and the unique-variable count saturates at $1715600$ for $N\geq28$. One copy of this state vector occupies $27.45\,\mathrm{MB}$. This is approximately $5.59$ times the storage of the fully symmetric vector, but it remains independent of $N$ beyond the saturation threshold and is still negligible compared with the conventional hierarchy. The localized and cavity-site superposition cases require identical structural memory because both retain one distinguished molecular label; their different populations and coherences alter only the initialized variable values, not the propagated dimension.

\section{Conclusions}
\label{sec:conclusions}

\noindent We have developed a symmetry adapted HEOM formulation for identical Holstein-Tavis-Cummings ensembles with independent, identical local environments. The reduction proceeds in two stages. Molecularly relabeled hierarchy elements are first represented by one canonical ADO occupation pattern. Repeated matrix elements within each representative are then replaced by a compact set of unique complex variables. These variables are propagated directly using precomputed hierarchy connections and molecular multiplicities, without constructing full ADO matrices or a dense HEOM generator.

The reduction is algebraically equivalent to the corresponding finite conventional HEOM defined by the same bath correlation decomposition, hierarchy depth, closure, and system-bath model. Canonicalization and unique-variable propagation introduce no additional dynamical approximation. The numerical comparison presented here shows that the reduced and conventional HEOM population dynamics are indistinguishable on the plotted scale.

For fixed hierarchy depth $L$ and number of exponential terms $m$, the number of canonical ADO representatives becomes independent of ensemble size for $N\geq L$, and the unique-variable dimension saturates for $N\geq L+2$. Fully permutation-symmetric initial operators use the maximally compressed representation. Localized populations and coherent nonsymmetric states are treated by retaining the required molecular labels explicitly while continuing to compress the remaining equivalent molecules. Linearity and molecular relabeling further permit general initial system operators, under the assumed initial bath preparation, to be reconstructed from a finite set of representative propagations.

The complete reduction requires identical molecular transition energies and cavity couplings, identical independent local environments, and local diagonal system-bath interactions. The present formulation is also restricted to two level molecules, one cavity mode, and the single excitation manifold. Moreover, although the propagated dimension saturates with $N$, its value can still grow rapidly with the number of bath correlation exponentials and hierarchy depth. Extensions to optical response functions, higher excitation manifolds, and ensembles of identical multichromophoric units will be considered separately in the forthcoming papers.
\\

\section*{Data availability}
\noindent The data that support the plots within this paper and other findings of this study are available from the corresponding authors upon a reasonable request.

\section*{Code availability}
\noindent The source code for this project are available from the corresponding authors upon a reasonable request. 
\\

\begin{acknowledgments}
This work was supported by the National Science Foundation Award under Grant No. CHE-2244683. M.E.M. appreciates the support from the Agnes M. and George Messersmith Fellowship by the University of Rochester. Computing resources were provided by the Center for Integrated Research Computing (CIRC) at the University of Rochester. We appreciate valuable suggestions from Sebastian Montillo Vega, Eric Koessler, and Santanu Poddar.\\
\end{acknowledgments}

\bibliographystyle{unsrtnat}
\bibliography{references}

\end{document}